\documentclass[final,12pt,times]{elsarticle}

\usepackage{graphicx}

\usepackage{tikz}
\usetikzlibrary{matrix,arrows}
\usetikzlibrary{automata,positioning,arrows}

\usepackage{pgfplots}
\usepackage{pgfplotstable}
\usepgfplotslibrary{fillbetween}
\pgfplotsset{compat=1.12}
\usepackage{booktabs}
\usepackage{caption}
\usepackage[font=normalsize]{subcaption}

\usepackage{nicefrac}
\usepackage{amsmath,amsthm,amsfonts}
\usepackage{mathtools}
\usepackage{bm}

\newcommand{\fref}[1]{Figure~\ref{#1}}
\newcommand{\sref}[1]{Section~\ref{#1}}
\newcommand{\vm}[1]{\bm{#1}}
\newcommand{\vx}{\vm{x}}

\newcommand{\compconj}[1]{%
  \overline{#1}%
}

\graphicspath{ {./figs/} } 
\DeclareGraphicsExtensions {.png,.pdf,.jpg}

\biboptions{sort&compress} 

\journal{Journal Name}

\begin{document}

\begin{frontmatter}

\title{Orbital-enriched flat-top partition of unity method for the
       Schr\"odinger eigenproblem}

\cortext[cor1]{Corresponding authors}
\author[label1]{Clelia Albrecht}
\author[label1]{Constanze Klaar}
\author[label2]{John Ernest Pask}
\author[label1,label3]{Marc Alexander Schweitzer\corref{cor1}}
\ead{schweitzer@ins.uni-bonn.de}
\author[label4]{N.~Sukumar\corref{cor1}}\ead{nsukumar@ucdavis.edu}
\author[label1]{Albert Ziegenhagel}
\address[label1]{Fraunhofer Institute for Algorithms and Scientific
                 Computing SCAI, Schloss Birlinghoven, 53757 Sankt 
                 Augustin, Germany}
\address[label2]{Physics Division, Lawrence Livermore National
                 Laboratory, 7000 East Avenue, Livermore, CA 94550, U.S.A.}
\address[label3]{Institute for Numerical Simulation, Rheinische 
                 Friedrich-Wilhelms-Universit{\"a}t, Wegelerstr. 6, 
                 53115 Bonn, Germany}
\address[label4]{Department of Civil and Environmental Engineering,
                 University of California, Davis, CA 95616, U.S.A.}
 
\begin{abstract}
Quantum mechanical calculations require the repeated solution of a Schr\"odinger equation for the wavefunctions of the system, from which materials properties follow. Recent work has shown the 
effectiveness of enriched finite element type Galerkin methods at significantly reducing the degrees of freedom required to obtain accurate solutions. However, time to solution has been 
adversely affected by the need to solve a generalized rather than standard eigenvalue problem and the ill-conditioning of associated systems matrices. In this work, we address both issues
by proposing a stable and efficient orbital-enriched partition
of unity method to solve the Schr\"odinger boundary-value problem in a
parallelepiped unit cell subject to Bloch-periodic boundary conditions.
In the proposed partition of unity method, the three-dimensional
domain is covered by
overlapping patches, with a compactly-supported weight function associated
with each patch. A key ingredient in our approach is the use of non-negative
weight functions that possess the {\it flat-top} property, i.e., each weight function
is identically equal to unity over some finite subset of its support. 
This flat-top property provides a pathway to devise a stable
approximation
over the whole domain. On each patch, we use
$p$-th degree orthogonal (Legendre) polynomials that ensure 
$p$-th order completeness, and in addition include eigenfunctions of the radial
 Schr\"odinger equation. Furthermore, we 
adopt a variational lumping approach to construct a (block-)diagonal overlap
matrix that yields a standard eigenvalue problem for which there
exist efficient eigensolvers.  The accuracy, stability, and efficiency
of the proposed method is demonstrated for the Schr\"odinger equation
with a harmonic potential as well as a localized Gaussian potential.
We show that the proposed approach delivers optimal rates of convergence in the
energy, and the use of orbital enrichment significantly reduces the
number of degrees of freedom for a given desired 
accuracy in the energy eigenvalues while the stability of the enriched
approach is fully maintained.
\end{abstract}
\end{frontmatter}

\section{Introduction}
The Kohn--Sham (KS) equations of density functional theory (DFT) are the dominant 
theoretical formulation in quantum mechanical simulations of condensed matter 
(solids and liquids). The KS equations require the
repeated solution of the steady-state Schr\"odinger and Poisson equations
on a parallelepiped unit cell with Bloch-boundary 
conditions~\cite{ashcroft:book}. The solution of the
Schr\"odinger equation is the most time-consuming part in KS-DFT calculations.
The current state-of-the-art
approach to solve the equations of KS-DFT is the planewave (PW)
pseudopotential method that uses a Fourier basis set, and requires the
solution of a discrete standard eigenproblem.
It has been appreciated in recent years that enriched Galerkin 
methods~\cite{Sukumar:Pask:2009,Lin:2012:ALB,Zhang:2017:ALB,Pask:Sukumar:2017,Yamakawa:2003,Yamakawa:2005,Chen:2007,Pask:Sukumar:Guney:Hu:2011,Pask:Sukumar:Mousavi:2012,Banerjee:2016,Kanungo:2017,Davydov:2016}
can be very competitive with PW methods in attaining the desired
accuracy with comparable or far fewer degrees of freedom (basis functions).
While early formulations~\cite{Yamakawa:2003,Yamakawa:2005,Sukumar:Pask:2009,Pask:Sukumar:Mousavi:2012}
employed direct enrichment, more recent approaches have employed discontinuous Galerkin~\cite{Lin:2012:ALB,Zhang:2017:ALB,Banerjee:2016} 
or partition of unity finite element formulations~\cite{Sukumar:Pask:2009,Pask:Sukumar:Guney:Hu:2011,Pask:Sukumar:2017,Davydov:2016}
in order to strictly localize orbital enrichments, thus facilitating flexible approximation and efficient 
parallel implementation. 
In~\cite{Pask:Sukumar:2017} it was shown that an enriched partition of unity finite element method (PUFEM)~\cite{BaMe:96,BaMe:97}
requires an order of magnitude fewer basis functions
than current state-of-the-art PW based methods to attain the desired $1$ mHa/atom accuracy in total energy 
calculations. However, the ill-conditioning of the resulting 
system matrices and the need to solve a generalized rather than standard eigenvalue problem
were key issues identified as adversely affecting time to solution in practice. In this work, we use
a {\it flat-top} partition of unity method (PUM)~\cite{MGriebel:MASchweitzer:2001a,MASchweitzer:2008d,Schweitzer:2013*2} 
to address these issues in the approximation of the Schr\"odinger
equation.  Our flat-top PUM produces well-conditioned system
(Hamiltonian and overlap) matrices and yields a standard
eigenvalue problem via variational lumping. The approximation quality of our flat-top PUM is comparable to that
reported in~\cite{Sukumar:Pask:2009}, but it overcomes 
the two main shortcomings of the PUFEM that arise in the solution
of the Schr\"odinger eigenproblem.
In addition to the electronic Schr\"odinger equation, the
flat-top PUM with Bloch boundary conditions also holds promise
in areas such as acoustic scattering~\cite{Ihlenburg:FEA:1998},
elastodynamics~\cite{Deymier:AMP:2013} and
electromagnetics~\cite{Joannopoulos:PCM:2008}, where large-scale 
eigenproblems are solved and useful {\em a priori\/} information is 
available.

In condensed matter calculations, the Schr\"odinger equation is 
solved in a unit cell (parallelepiped domain $\Omega$)
subject to Bloch-periodic boundary conditions (see Figure~\ref{figure:omega}). In the flat-top
partition of unity method, the domain $\Omega$ is covered
by overlapping patches (see Figure~\ref{figure:uniformCover}) and each patch $i$ is associated with
a weight function $\varphi_i(\vx)$ with support $\omega_i$ such that
$\sum_i \varphi_i(\vx) = 1$ and $\varphi_i(\vx) \equiv 1$ on $\omega_i^{\textrm{FT}} \subset
\omega_i$ (see Figure~\ref{figure:comparePU}). The local basis set $V_i$ on each patch consists 
of polynomials and/or non-polynomial
(orbital enrichment) functions, and the global approximation
is formed by linear combinations of the products of $\varphi_i(\vx)$ and functions from $V_i$.
We perform local orthogonalization to ensure that all functions on a patch
are linearly independent and thereby attain global stability~\cite{MASchweitzer:2008d},
and adopt the variational lumping
scheme~\cite{Schweitzer:2013*2} to realize a standard eigenproblem.

The remainder of the paper is organized as follows. In the next section,
we state the strong and weak forms of the Schr{\"o}dinger
eigenproblem. In~\sref{section:partition_of_unity_method}, we introduce the 
partition of unity method, where we present the proposed flat-top PUM in~\sref{subsection:flat_top_PUM}.
The key steps in the local orthogonalization procedure to construct a
stable global approximation are discussed 
in~\sref{section:orbital_enrichment_and_stability}, and we describe the
variational lumping scheme in~\sref{section:variational_mass_lumping}.  Numerical
examples for the Schr{\"o}dinger equation are presented in
\sref{section:numerical_results}, where we show that the system matrices
are well-conditioned and that the use of orbital-enrichment provides a very
efficient solution vis-{\`a}-vis solely using polynomials over
each patch. In addition, we also provide comparisons in the
eigenspectrum when using the consistent overlap matrix versus the 
lumped overlap matrix, and the results reveal that the
variational lumping scheme does not adversely affect the 
accuracy of the energy eigenvalues. We close with a few
concluding remarks in~\sref{section:concluding_remarks}.

\section{The Schr\"odinger eigenproblem}
\label{section:kohn_sham_equations}
\begin{figure}[tbp]
  \centering
  \scalebox{1}{
\begin{tikzpicture}[every node/.style={minimum size=1cm},on grid,scale=0.8]
\begin{scope}[every node/.append style={yslant=-0.5},yslant=-0.5]
  \shade[right color=gray!10, left color=black!50] (0,0) rectangle +(3,3);

  \node at (1.5,1.5) {$\Gamma_3$};
  \node at (2.7,1.5) {$\vm{a}_1$};
  \draw[->, ultra thick](2.999,3)--(2.999,0);

\end{scope}
\begin{scope}[every node/.append style={yslant=0.5},yslant=0.5]
  \shade[right color=gray!70,left color=gray!10] (3,-3) rectangle +(3,3);
   \node at (4.5,-0.3) {$\vm{a}_3$};
  \node at (4.5,-1.5) {$\Gamma_2$};
   \draw[->,ultra thick] (3,0)--(6,0);
\draw[->,ultra thick] (3,0)--(3,-3);
\end{scope}
\begin{scope}[every node/.append style={
    yslant=0.5,xslant=-1},yslant=0.5,xslant=-1
  ]
  \shade[bottom color=gray!10, top color=black!80] (6,3) rectangle +(-3,-3);
   \node at (3.3,1.5) {$\vm{a}_2$};
  \node at (4.5,1.5) {$\Gamma_1$};
\draw[->, ultra thick](3,0)--(6,0);
\draw[->,ultra thick](3,0)--(3,3);
\end{scope}
\end{tikzpicture}}
\caption{Sketch of a parallelepiped unit cell $\Omega$ spanned by primitive lattice vectors $\vm{a}_d$ with respective boundary segments $\Gamma_d$ ($d=1,2,3$).}
\label{figure:omega}
\end{figure}
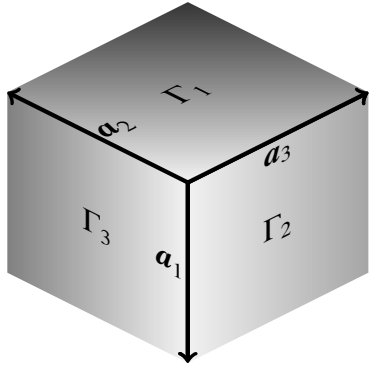

The stationary Schr\"odinger equation reads as 
\begin{equation}
\label{equation:schroedinger_equation_ks_orbital}
\mathcal{H} \psi(\vm{x}) \coloneqq - \frac{1}{2} \nabla^2 \psi(\vm{x}) + {V}_{\mathrm{eff}}(\vm{x})\psi(\vm{x}) = \varepsilon \psi(\vm{x}) \hspace{1em}\mbox{ in } \Omega.
\end{equation}
We consider a parallelepiped unit-cell $\Omega \subset \mathbb{R}^3$ with primitive lattice vectors $\vm{a}_d$ ($d=1,2,3$) to describe a periodic condensed matter system (see Figure 
\ref{figure:omega}). Thus, the effective potential and the charge density are periodic, i.e., they satisfy
\begin{equation}
\label{equation:periodicity_density}
{V}_{\mathrm{eff}}(\vm{x} + \vm{R}) = {V}_{\mathrm{eff}}(\vm{x}), \quad \rho(\vm{x} + \vm{R}) = \rho(\vm{x}),
\end{equation}
whereas the solution of Schr{\"o}dinger's equation $\psi$, the so-called wavefunction, satisfies Bloch's theorem
\begin{equation}
\label{equation:periodicity_wavefunction}
\psi(\vm{x} + \vm{R}) = \psi(\vm{x}) \exp(i\vm{k}\cdot\vm{R})
\end{equation}
for any lattice translation vector $\vm{R}=n_1\vm{a}_1+n_2\vm{a}_2+n_3\vm{a}_3$ with $n_d \in \mathbb{Z}$ ($d=1,2,3$) and wavevector $\vm{k}$. 
Note that for $\vm{k}=0$ ($\Gamma$-point) the wavefunction is also periodic whereas for all other wavevectors there is a phase-shift $\exp(i\vm{k}\cdot\vm{R})$ 
associated with a translation by $\vm{R}$. In this situation the general problem~\eqref{equation:schroedinger_equation_ks_orbital} becomes 
\begin{equation}
\label{equation:schroedinger_strong}
\begin{array}{rclll}
- \frac{1}{2} \nabla^2 \psi(\vm{x}) + {V}_{\mathrm{eff}}(\vm{x})\psi(\vm{x}) & = & \varepsilon \psi(\vm{x}) & \mbox{ in } \Omega, \\
\psi(\vm{x}+\vm{a}_d) & = & \exp(i\vm{k} \cdot \vm{a}_d) \psi(\vm{x}) & \mbox{ on } \Gamma_d, \\
\nabla\psi(\vm{x}+\vm{a}_d) \cdot \hat{\vm{n}}(\vm{x}) & = & \exp(i\vm{k} \cdot \vm{a}_d) \nabla\psi(\vm{x}) \cdot \hat{\vm{n}}(\vm{x}) & \mbox{ on } \Gamma_d,
\end{array}
\end{equation}
where $(\psi, \varepsilon)$ denotes an eigenpair consisting of the respective wavefunction $\psi$ and its associated energy $\varepsilon$, $\hat{\vm{n}}(\vm{x})$ is the  
outward unit normal at $\vm{x}$ and $\Gamma_d$ are the bounding faces of the domain $\Omega$, see~\fref{figure:omega}. 
Even though the boundary conditions and thus the wavefunction are complex-valued, the eigenvalues $\varepsilon \in \mathbb{R}$ due to the self-adjointness of the Hamiltonian $\mathcal{H}$.

In this paper, we are concerned with the numerical approximation of~\eqref{equation:schroedinger_strong} by Galerkin methods and thus need to rewrite~\eqref{equation:schroedinger_strong} 
in its respective weak form. To this end, we consider the function space 
\begin{equation}
\label{equation:function_space_weak_form}
\mathcal{V} \coloneqq \{v \in H^1(\Omega, \mathbb{C}) : v(\vm{x} + \vm{a}_d) = v(\vm{x}) \exp(i\vm{k} \cdot \vm{a}_d) \text{ on } \Gamma_d, \, d=1,2,3 \}
\end{equation}
and test~\eqref{equation:schroedinger_strong} with $v \in \mathcal{V}$ to attain
\begin{equation}
\label{equation:weak_form_eigenvalue_problem}
a(v, \psi) = \varepsilon \langle v, \psi\rangle_{L^2(\Omega,\mathbb{C})} \text{ for all } v \in \mathcal{V}
\end{equation}
with 
\begin{equation}
\label{equation:weak_form_hamiltonian_bilinearform}
a(v, \psi) \coloneqq \frac{1}{2} \int_{\Omega} \Bigl(\compconj{\nabla v(\vm{x})}\nabla \psi(\vm{x}) + \compconj{v(\vm{x})} {V}_{\mathrm{eff}}(\vm{x})\psi(\vm{x})\Bigr) \, d\vm{x}
\end{equation}
and
\begin{equation}
\label{equation:weak_form_mass_bilinearform}
\langle v, \psi\rangle_{L^2(\Omega,\mathbb{C})} \coloneqq \int_{\Omega} \compconj{v(\vm{x})}\psi(\vm{x}) \, d\vm{x}
\end{equation}
after integration by parts~\cite{Sukumar:Pask:2009}. 
Thus, choosing a finite-dimensional space $V_M \subset \mathcal{V}$ with basis functions $\phi_i$ to discretize~\eqref{equation:weak_form_eigenvalue_problem} we obtain the generalized eigenproblem
\begin{equation}
\label{equation:generalized_eigenvalue_problem}
H \tilde{\psi} = \varepsilon S \tilde{\psi} ,
\end{equation}
where $H=(H_{ij})\in \mathbb{C}^{M \times M}$ denotes the discrete Hamiltonian and $S=(S_{ij})\in \mathbb{C}^{M \times M}$ is the so-called overlap (or consistent mass) matrix 
\begin{equation}
H_{ij} \coloneqq a(\phi_j, \phi_i), \text{ and } S_{ij} \coloneqq \langle \phi_j, \phi_i \rangle_{L^2(\Omega,\mathbb{C})}.
\end{equation}
The approximate eigenfunction $\psi \in V_M$ is given by
\begin{equation}
\psi(\vm{x}) \coloneqq \sum_{i=1}^M \psi_i \phi_i(\vm{x}),
\end{equation}
where $\tilde{\psi} = (\psi_i) \in \mathbb{C}^M$ denotes the associated coefficient vector. Throughout this paper we employ a particular partition of unity method~\cite{MGriebel:MASchweitzer:2001a,MASchweitzer:2003,Klaar:2016} to construct the respective finite-dimensional 
space $V_M$ and its basis functions $\phi_i$.

\section{Partition of unity method}
\label{section:partition_of_unity_method}
The partition of unity method (PUM) is a generalization of the finite element method (FEM) that is typically employed for the spatial discretization of a partial differential equation (PDE), 
see for example~\cite{Fries:Belytschko:2010,Schweitzer:2012}. The notion of a PUM was coined in \cite{BaMe:96,BaMe:97} and is based on the special FEM developed in~\cite{Babuska:Caloz:Osborn:1994}. 
The abstract ingredients of a PUM are a partition of unity (PU) $\{\varphi_i:i=1,\ldots,N\}$ and a collection of local approximation spaces 
$V_i(\omega_i) \coloneqq \operatorname{span} \langle \vartheta_i^m\rangle_{m=1}^{\operatorname{dim}(V_i)}$ defined on the patches $\omega_i\coloneqq\operatorname{supp}(\varphi_i)$ for $i=1,\ldots,N$ which overlap and whose 
union covers the computational domain $\Omega \subset \mathbb{R}^d$. With these two ingredients we define the PUM space
\begin{equation}
\label{equation:pumSpaceDefinition}
V^{\mathrm{PU}} \coloneqq \sum_{i=1}^N \varphi_i V_i = \operatorname{span} \langle \varphi_i \vartheta_i^m\rangle ;
\end{equation}
that is, the basis functions of a PUM space are simply defined as the products of the PU functions $\varphi_i$ and the local approximation functions $\vartheta_i^m$. The PU functions provide 
the locality and global regularity of the product functions whereas the functions $\vartheta_i^m$ equip $V^{\mathrm{PU}}$ with its approximation power. Note that the local approximation 
spaces $V_i$ can be chosen in a problem-dependent fashion and are independent of each other, i.e., any local basis $\vartheta_i^m$ may be employed on any patch $\omega_i$. 
In general the local approximation space $V_i \coloneqq \operatorname{span} \langle \vartheta_i^n \rangle$ associated
with a particular patch $\omega_i$ of a PUM space $V^{\mathrm{PU}}$ consists of two parts: A smooth approximation
space, for example polynomials $V^\mathcal{P}_i (\omega_i) \coloneqq \operatorname{span} \langle \pi_i^s \rangle$, and a problem-dependent
enrichment part $V^\mathcal{E}_i(\omega_i) \coloneqq \operatorname{span} \langle \psi_i^t \rangle$, i.e.,
\begin{equation}
\label{equation:localApproximationSpaceDefinition}
\operatorname{span} \langle \vartheta_i^n \rangle = V_i (\omega_i) = V^\mathcal{P}_i (\omega_i) + V^\mathcal{E}_i(\omega_i) = \operatorname{span} \langle \pi_i^s, \psi_i^t \rangle.
\end{equation}

For the ease of notation, we make the following conventions. For an arbitrary function $u \in V^{\mathrm{PU}}$ with the representation
\begin{displaymath}
u(x) = \sum_{i=1}^N \sum_{m=1}^{\operatorname{dim}(V_i)} u_i^m \varphi_i (x) \vartheta_i^m (x) = \sum_{i=1}^N \varphi_i (x) \sum_{m=1}^{\operatorname{dim}(V_i)} u_i^m \vartheta_i^m (x) =: \sum_{i=1}^N \varphi_i (x) u_i(x)
\end{displaymath}
we denote its associated overall coefficient vector by
\begin{displaymath}
\tilde{u} = (u_{(i,m)}) \in \mathbb{R}^{\operatorname{dim}(V^\mathrm{PU})} \text{ with } \operatorname{dim}(V^\mathrm{PU}) \coloneqq\sum_{i=1}^N \operatorname{dim}(V_i).
\end{displaymath}

\subsection{Flat-top partition of unity}
\label{subsection:flat_top_PUM}
\begin{figure}[tbp]
  \includegraphics[width=0.32\textwidth] {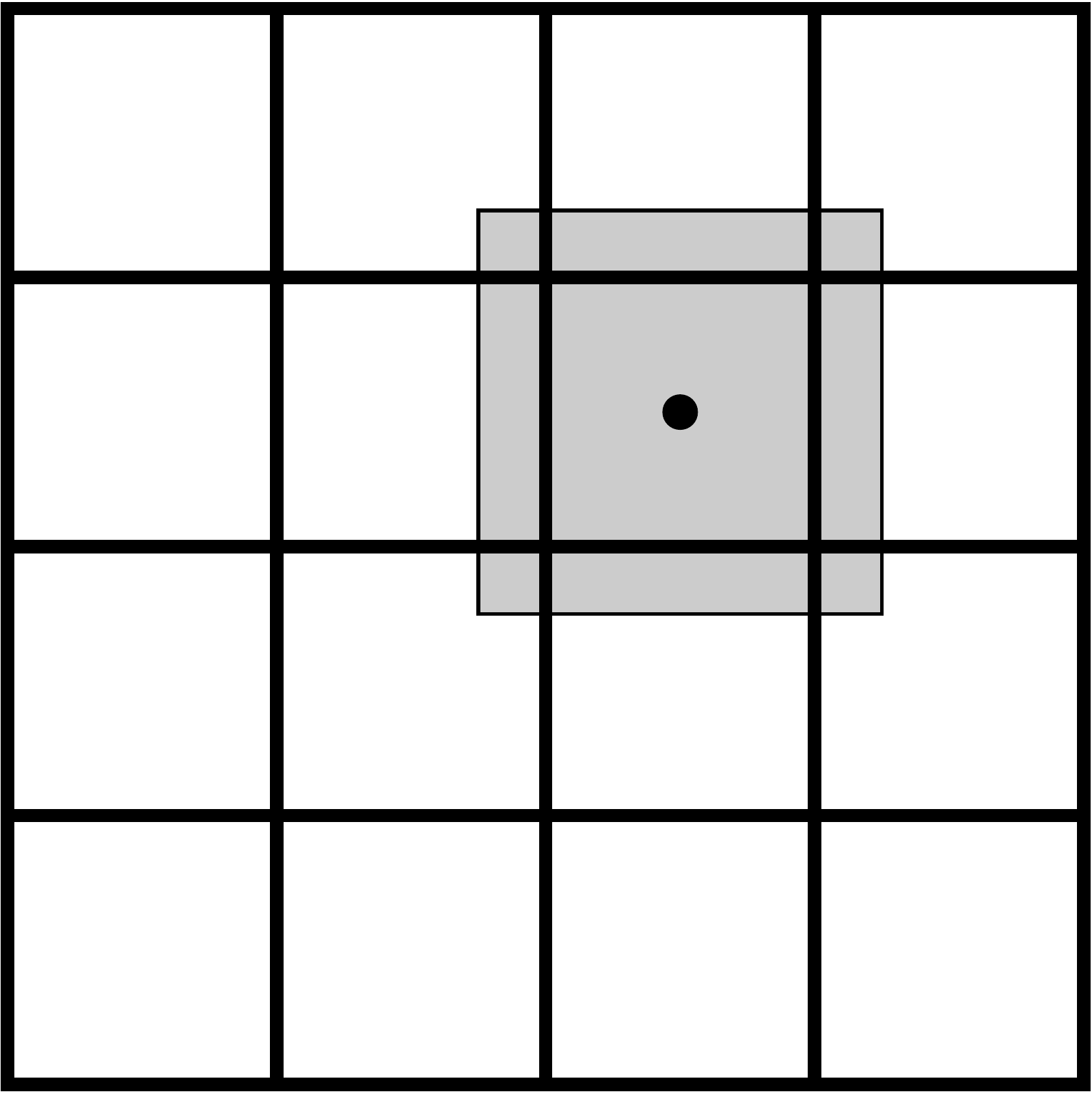}\hfill
  \includegraphics[width=0.32\textwidth] {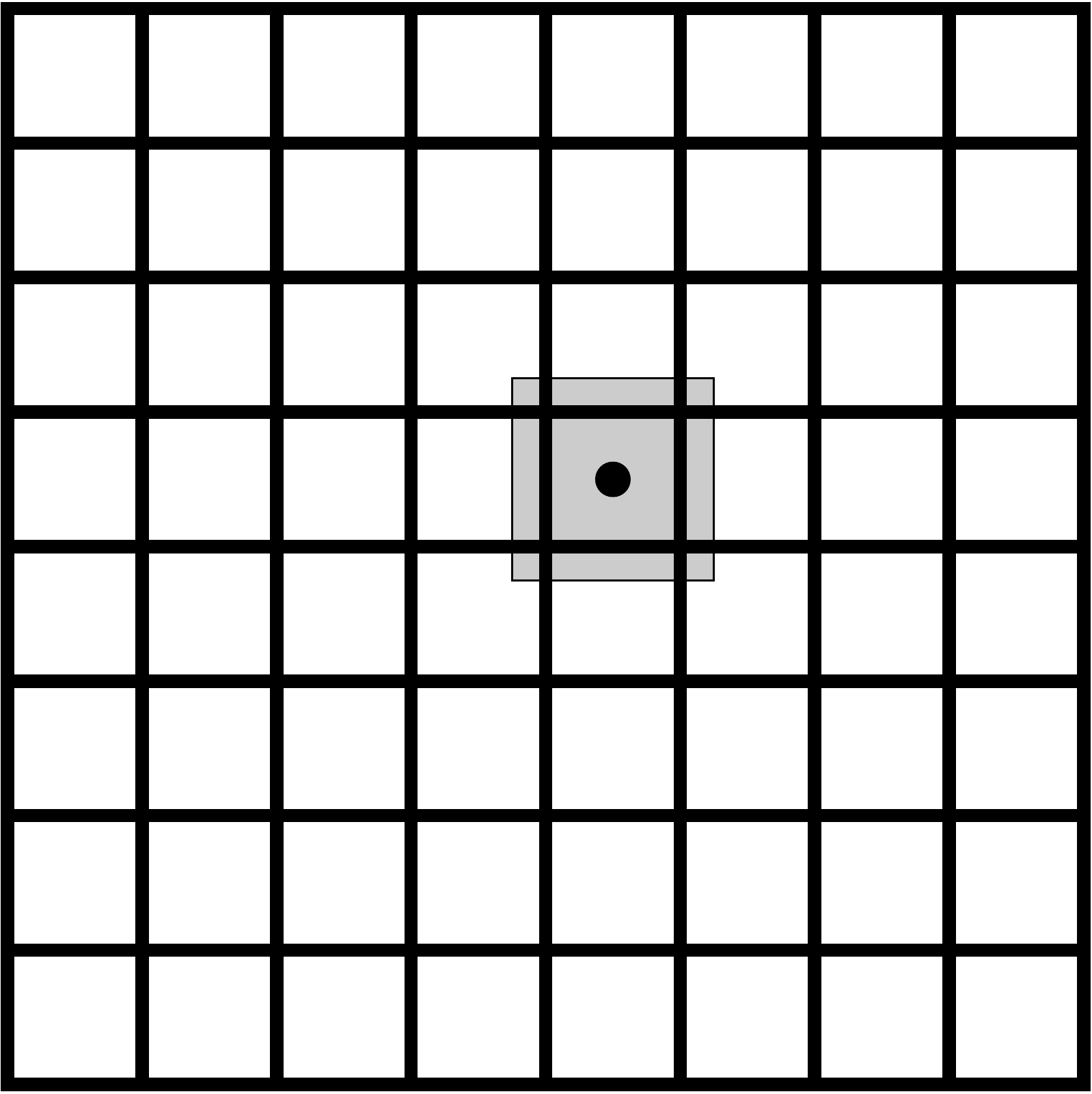}\hfill
  \includegraphics[width=0.32\textwidth] {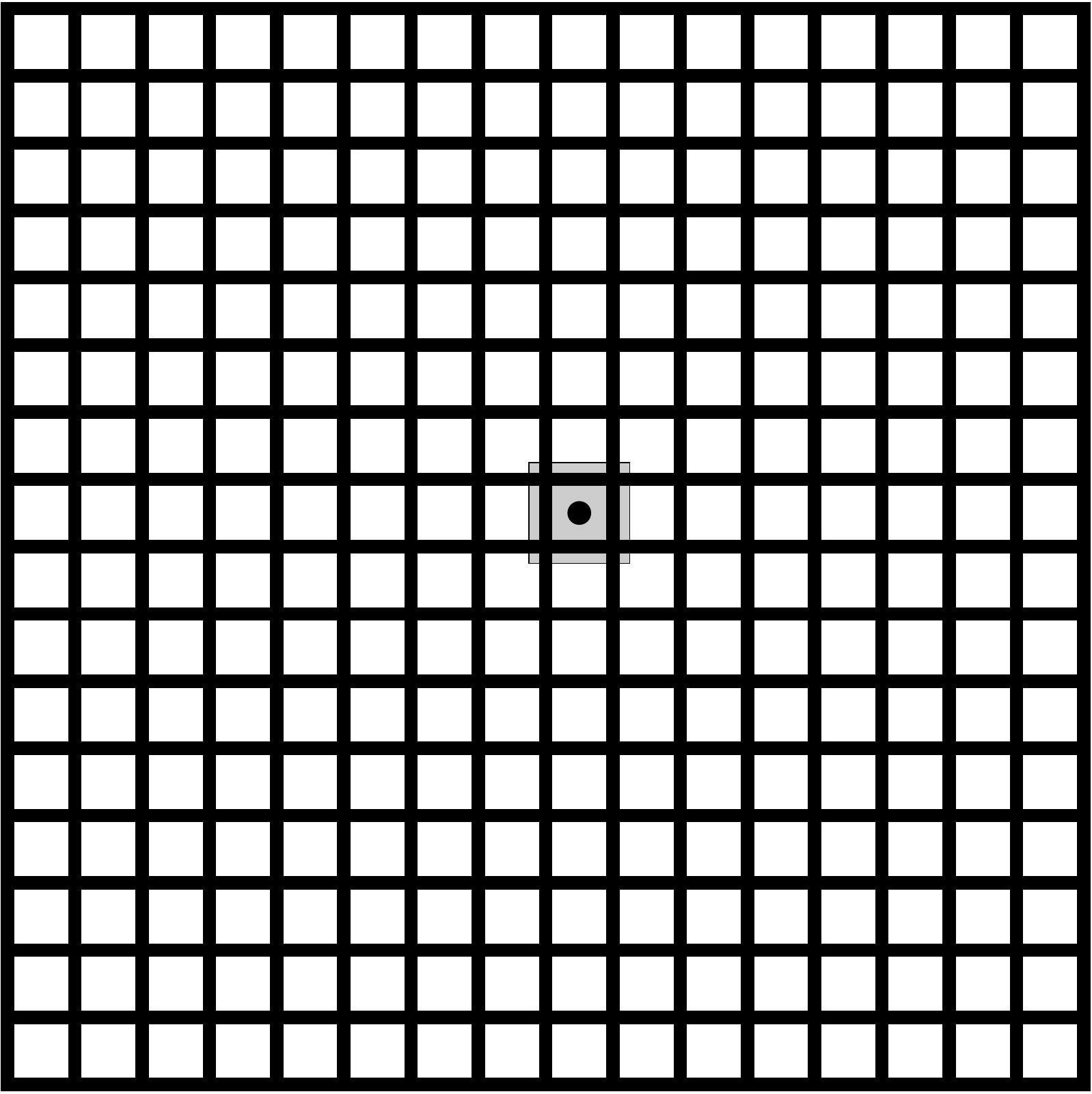}\\
  \caption{Schematic of a sequence of uniformly refined covers which come from the scaling of uniform grid cells in two dimensions. Depicted is a single cover patch 
  $\omega_i=\prod_{l=1}^d (o_i^l-\alpha h, o_i^l+\alpha h)$ (gray) with $2h=\nicefrac{1}{4},\nicefrac{1}{8},\nicefrac{1}{16}$ (left to right) and its center $o_i$.}
\label{figure:uniformCover}
\end{figure}
\begin{figure}[tbp]
\centering
\scalebox{0.85}{
\begin{tikzpicture}
\begin{axis}[
    xmin=-0.1,
    xmax=1.1,
    ymin=-0.100,
    ymax=1.10,
    ytick={0,1},
    xtick={0,0.25,0.5,0.75,1.0},
        every axis/.append style={font=\normalsize},
      axis y line=left,
  axis x line=middle,
]
\addplot[blue,domain=-0.5:0.5, no markers,samples=100,line width=1.5pt] coordinates {(0,1.0)    (0.5,0.0)};
\addplot[red,domain=0:1.0, no markers,samples=100,line width=1.5pt] coordinates {(0,0)   (0.5,1.0)    (1,0.0)};
\addplot[green,domain=0.5:1.5, no markers,samples=100,line width=1.5pt] coordinates {(0.5,0)   (1,1.0)};
\addplot[thick, samples=50, smooth,domain=0:6,gray, name path=three,dashed] coordinates {(0,0)(0,3)};          
\addplot[thick, samples=50, smooth,domain=0:6,gray, name path=three,dashed] coordinates {(1,0)(1,3)};
\end{axis} 
\end{tikzpicture}}
\scalebox{0.85}{
\begin{tikzpicture}
\begin{axis}[
    xmin=-0.1,
    xmax=1.1,
    ymin=-0.100,
    ymax=1.10,
    ytick={0,1},
    xtick={0,0.25,0.5,0.75,1.0},
        every axis/.append style={font=\normalsize},
      axis y line=left,
  axis x line=middle,
]
\addplot[blue,domain=-0.5:0.5, no markers,samples=100,line width=1.5pt] coordinates {(0,1.0)   (0.25,1.0)    (0.3755,0.0)};
\addplot[red,domain=0:1.0, no markers,samples=100,line width=1.5pt] coordinates {(0.25,0)   (0.375,1.0)    (0.625,1.0)    (0.75,0.0)};
\addplot[green,domain=0.5:1.5, no markers,samples=100,line width=1.5pt] coordinates {(0.625,0)   (0.75,1.0)    (1.0,1.0)};
\addplot[thick, samples=50, smooth,domain=0:6,gray, name path=three,dashed] coordinates {(0,0)(0,3)};          
\addplot[thick, samples=50, smooth,domain=0:6,gray, name path=three,dashed] coordinates {(1,0)(1,3)};
\end{axis} 
\end{tikzpicture}}\\
  \caption{Partition of unity comprised of linear FE functions (left) and a flat-top PU (right) in one dimension.}
\label{figure:comparePU}
\end{figure}
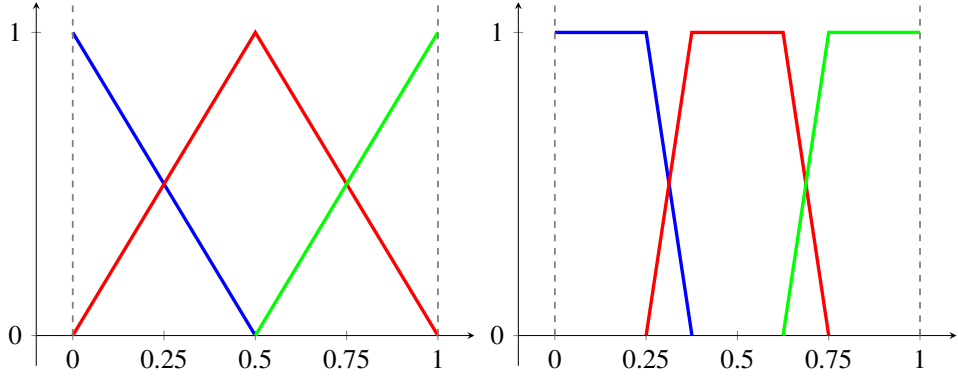

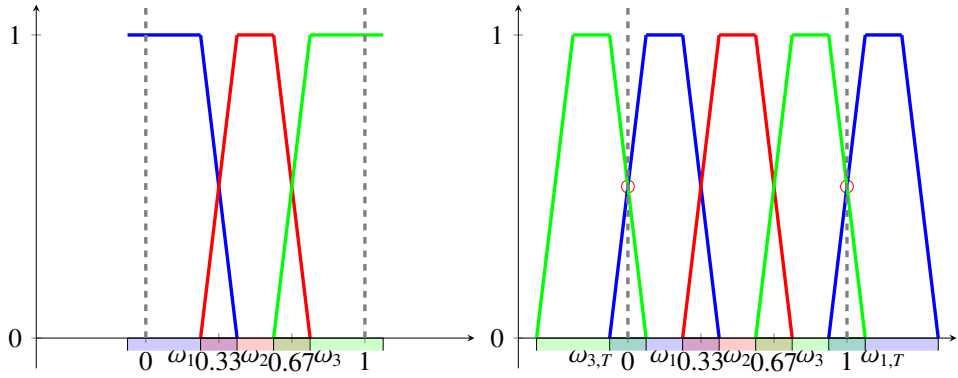
\begin{figure}[tbp]
 \centering
 \scalebox{0.85}{
\begin{tikzpicture}
\begin{axis}[
    xmin=-0.5,
    xmax=1.5,
    ymin=-0.100,
    ymax=1.10,
    ytick={0,1},
    xtick={0,0.333333333,0.6666666667,1.0},
        every axis/.append style={font=\normalsize},
    	axis y line=left,
	axis x line=middle,
]

 \addplot[blue,domain=-0.0833333333:0.25, no markers,samples=100,line width=1.5pt] {1}; 
 \addplot[blue,domain=0.25:0.41666666667, no markers,samples=100,line width=1.5pt] {-6*x+2.5};

 \addplot[red,domain=0.25:0.41666666667, no markers,samples=100,line width=1.5pt] {6*x-1.5}; 
 \addplot[red,domain=0.41666666667:0.5833333333, no markers,samples=100,line width=1.5pt] {1};
 \addplot[red,domain=0.5833333333:0.75, no markers,samples=100,line width=1.5pt] {-6*x+4.5}; 
 
 \addplot[green,domain=0.5833333333:0.75, no markers,samples=100,line width=1.5pt] {6*x-3.5}; 
 \addplot[green,domain=0.75:1.0833333333, no markers,samples=100,line width=1.5pt] {1};

                     \node[below] at(axis cs:0.166667,-0.009){$\omega_1$};
      
            \node[below] at(axis cs:0.5,-0.009){$\omega_2$};
            \node[below] at(axis cs:0.833333,-0.009){$\omega_3$};
  
  \addplot [fill=blue, fill opacity=0.2]coordinates {
         (-0.08333, -0.0415) 
         (-0.08333, 0)            
         (0.41666666667, 0)  
         (0.41666666667, -0.0415)};

\addplot [fill=red, fill opacity=0.2]coordinates {
         (0.25, -0.0415) 
         (0.25, 0)            
         (0.75, 0)  
         (0.75, -0.0415)};
\addplot [fill=green, fill opacity=0.2]coordinates {
         (0.5833333333, -0.0415) 
         (0.5833333333, 0)            
         (1.0833333333, 0)  
         (1.0833333333, -0.0415)};

   \addplot[line width=1.5pt, samples=50, smooth,domain=0:6,gray, name path=three,dashed] coordinates {(0,0)(0,3)};          
  \addplot[line width=1.5pt, samples=50, smooth,domain=0:6,gray, name path=three,dashed] coordinates {(1,0)(1,3)};
   \end{axis} 
\end{tikzpicture}}
\scalebox{0.85}{
\begin{tikzpicture}
\begin{axis}[
    xmin=-0.5,
    xmax=1.5,
    ymin=-0.100,
    ymax=1.10,
    ytick={0,1},
    xtick={0,0.333333333,0.6666666667,1.0},
        every axis/.append style={font=\normalsize},
    	axis y line=left,
	axis x line=middle,
]
     \draw[red] (axis cs:0,0.5) circle (1mm);
     \draw[red] (axis cs:1,0.5) circle (1mm);
 \addplot[blue,domain=-0.0833333333:0.0833333333, no markers,samples=100,line width=1.5pt] {6*x+0.5};
 \addplot[blue,domain=0.0833333333:0.25, no markers,samples=100,line width=1.5pt] {1}; 
 \addplot[blue,domain=0.25:0.41666666667, no markers,samples=100,line width=1.5pt] {-6*x+2.5};
 
  \addplot[blue,domain=0.9166666667:1.0833333333, no markers,samples=100,line width=1.5pt] {6*(x-1)+0.5};
 \addplot[blue,domain=1.0833333333:1.25, no markers,samples=100,line width=1.5pt] {1}; 
 \addplot[blue,domain=1.25:1.41666666667, no markers,samples=100,line width=1.5pt] {-6*(x-1)+2.5};
 
 \addplot[red,domain=0.25:0.41666666667, no markers,samples=100,line width=1.5pt] {6*x-1.5}; 
 \addplot[red,domain=0.41666666667:0.5833333333, no markers,samples=100,line width=1.5pt] {1};
 \addplot[red,domain=0.5833333333:0.75, no markers,samples=100,line width=1.5pt] {-6*x+4.5}; 
 
 \addplot[green,domain=0.5833333333:0.75, no markers,samples=100,line width=1.5pt] {6*x-3.5}; 
 \addplot[green,domain=0.75:0.9166666667, no markers,samples=100,line width=1.5pt] {1};
  \addplot[green,domain=0.9166666667:1.0833333333, no markers,samples=100,line width=1.5pt] {-6*x+6.5};

   \addplot[green,domain=-0.4166666667:-0.25, no markers,samples=100,line width=1.5pt] {6*(x+1)-3.5}; 
 \addplot[green,domain=-0.25:-0.0833333333, no markers,samples=100,line width=1.5pt] {1};
    \addplot[green,domain=-0.08333:0.0833333333, no markers,samples=100,line width=1.5pt] {-6*(x+1)+6.5};
  
         
            \node[below] at(axis cs:0.166667,-0.009){$\omega_1$};
            \node[below] at(axis cs:1.166667,-0.009){$\omega_{1,T}$};
            \node[below] at(axis cs:0.5,-0.009){$\omega_2$};
            \node[below] at(axis cs:0.833333,-0.009){$\omega_3$};
            \node[below] at(axis cs:-0.166667,-0.009){$\omega_{3,T}$};
  \addplot [fill=blue, fill opacity=0.2]coordinates {
         (-0.08333, -0.0415) 
         (-0.08333, 0)            
         (0.41666666667, 0)  
         (0.41666666667, -0.0415)};
           \addplot [fill=blue, fill opacity=0.2]coordinates {
         (0.91667, -0.0415) 
         (0.91667, 0)            
         (1.41666666667, 0)  
         (1.41666666667, -0.0415)};
\addplot [fill=red, fill opacity=0.2]coordinates {
         (0.25, -0.0415) 
         (0.25, 0)            
         (0.75, 0)  
         (0.75, -0.0415)};
\addplot [fill=green, fill opacity=0.2]coordinates {
         (0.5833333333, -0.0415) 
         (0.5833333333, 0)            
         (1.0833333333, 0)  
         (1.0833333333, -0.0415)};
 \addplot [fill=green, fill opacity=0.2]coordinates {
         (-0.4166666667, -0.0415) 
         (-0.4166666667, 0)            
         (0.0833333333, 0)  
         (0.0833333333, -0.0415)};
         
   \addplot[line width=1.5pt, samples=50, smooth,domain=0:6,gray, name path=three,dashed] coordinates {(0,0)(0,3)};          
  \addplot[line width=1.5pt, samples=50, smooth,domain=0:6,gray, name path=three,dashed] coordinates {(1,0)(1,3)};
   \end{axis} 
\end{tikzpicture}}
 \caption{Comparison of one-dimensional flat-top PU weight functions for Dirichlet or Neumann boundary conditions (left) and periodic boundary conditions (right). To realize periodic weight 
 functions we copy patches $\omega_i$ with $\omega_i \cap \partial \Omega \not=\emptyset$ periodically (see also~\fref{figure:periodic_neighbors} and~\cite{Klaar:2016}).}
\label{figure:periodic_neighbors_1D}
\end{figure}

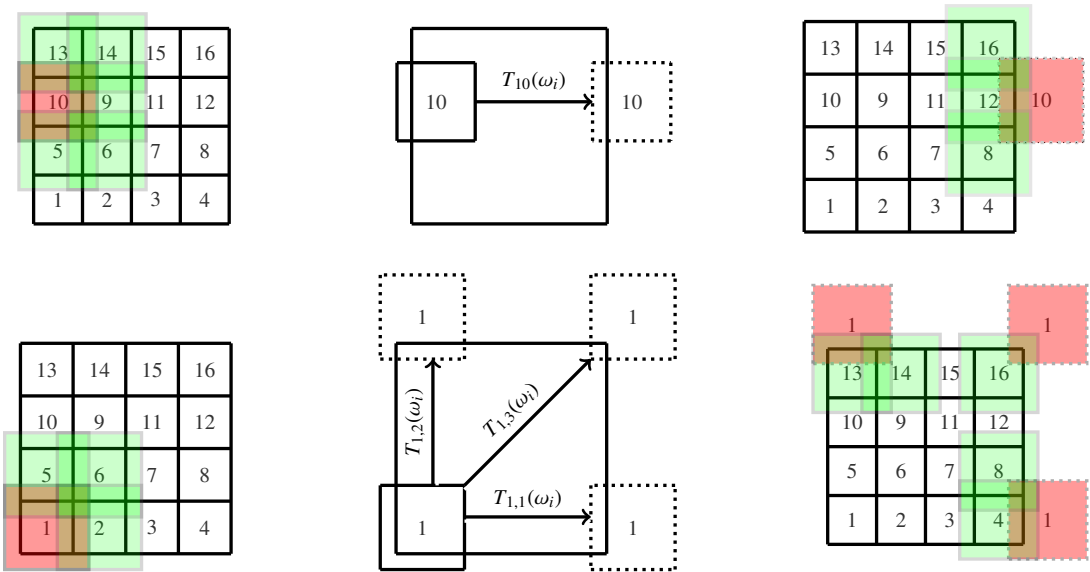
\begin{figure}
\centering
\begin{subfigure}{.33\textwidth}
\centering
\begin{tikzpicture}[scale=1.3]
\draw[-,line width=1.25pt](0,0)--(0,2);
\draw[-,line width=1.25pt](0,2)--(2,2);
\draw[-,line width=1.25pt](2,2)--(2,0);
\draw[-,line width=1.25pt](0,0)--(2,0);

\draw[-,line width=1.25pt](0.5,0)--(0.5,2);
\draw[-,line width=1.25pt](1,0)--(1,2);
\draw[-,line width=1.25pt](1.5,0)--(1.5,2);

\draw[-,line width=1.25pt](0,0.5)--(2,0.5);
\draw[-,line width=1.25pt](0,1)--(2,1);
\draw[-,line width=1.25pt](0,1.5)--(2,1.5);

\node at (0.25,0.25) {\textcolor {black!80}{\scriptsize 1}};
\node at (0.75,0.25) {\textcolor {black!80}{\scriptsize 2}};
\node at (1.25,0.25) {\textcolor {black!80}{\scriptsize 3}};
\node at (1.75,0.25) {\textcolor {black!80}{\scriptsize 4}};
\node at (0.25,0.75) {\textcolor {black!80}{\scriptsize 5}};
\node at (0.75,0.75) {\textcolor {black!80}{\scriptsize 6}};
\node at (1.25,0.75) {\textcolor {black!80}{\scriptsize 7}};
\node at (1.75,0.75) {\textcolor {black!80}{\scriptsize 8}};
\node at (0.75,1.25) {\textcolor {black!80}{\scriptsize 9}};
\node at (0.25,1.25) {\textcolor {black!80}{\scriptsize 10}};
\node at (1.25,1.25) {\textcolor {black!80}{\scriptsize 11}};
\node at (1.75,1.25) {\textcolor {black!80}{\scriptsize 12}};
\node at (0.25,1.75) {\textcolor {black!80}{\scriptsize 13}};
\node at (0.75,1.75) {\textcolor {black!80}{\scriptsize 14}};
\node at (1.25,1.75) {\textcolor {black!80}{\scriptsize 15}};
\node at (1.75,1.75) {\textcolor {black!80}{\scriptsize 16}};

\draw [fill=red,opacity=.4,draw=black!80,line width=1.25pt ] (-0.15,0.85) rectangle (0.65,1.65);

\draw [fill=green,opacity=.2,draw=black!80,line width=1.25pt ] (-0.15,0.35) rectangle (0.65,1.15);
\draw [fill=green,opacity=.2,draw=black!80,line width=1.25pt ] (-0.15,1.35) rectangle (0.65,2.15);
\draw [fill=green,opacity=.2,draw=black!80,line width=1.25pt ] (0.35,1.35) rectangle (1.15,2.15);
\draw [fill=green,opacity=.2,draw=black!80,line width=1.25pt ] (0.35,0.85) rectangle (1.15,1.65);
\draw [fill=green,opacity=.2,draw=black!80,line width=1.25pt ] (0.35,0.35) rectangle (1.15,1.15);
\end{tikzpicture}

\end{subfigure}%
\begin{subfigure}{.33\textwidth}
\centering
\begin{tikzpicture}[scale=1.3]
   \draw[-,line width=1.25pt ](0,0)--(0,2);
\draw[-,line width=1.25pt ](0,2)--(2,2);
\draw[-,line width=1.25pt ](2,2)--(2,0);
\draw[-,line width=1.25pt ](0,0)--(2,0);

\draw[-,line width=1.25pt ](-0.15,0.85)--(-0.15,1.65);
\draw[-,line width=1.25pt ](-0.15,1.65)--(0.65,1.65);
\draw[-,line width=1.25pt ](0.65,1.65)--(0.65,0.85);
\draw[-,line width=1.25pt ](0.65,0.85)--(-0.15,0.85);

\draw[dotted,line width=1.25pt ](1.85,0.85)--(1.85,1.65);
\draw[dotted,line width=1.25pt ](1.85,1.65)--(2.65,1.65);
\draw[dotted,line width=1.25pt ](2.65,1.65)--(2.65,0.85);
\draw[dotted,line width=1.25pt ](2.65,0.85)--(1.85,0.85);
\draw[fill=green,opacity=.0,draw=black!00,line width=1.25pt ](2.64,2.14)--(2.65,2.15);

\node at (0.25,1.25) {\textcolor {black!80}{\scriptsize 10}};
\node at (2.25,1.25) {\textcolor {black!80}{\scriptsize 10}};

\draw[->,line width=1.25pt](0.65,1.25)--node[above]{\scriptsize $T_{10}(\omega_i)$}(1.85,1.25);
\end{tikzpicture}
\end{subfigure}%
\begin{subfigure}{.33\textwidth}
\centering
\begin{tikzpicture}[scale=1.4]

\draw[dotted ](1.85,0.85)--(1.85,1.65);
\draw[dotted ](1.85,1.65)--(2.65,1.65);
\draw[dotted ](2.65,1.65)--(2.65,0.85);
\draw[dotted ](2.65,0.85)--(1.85,0.85);

\node at (0.25,0.25) {\textcolor {black!80}{\scriptsize 1}};
\node at (0.75,0.25) {\textcolor {black!80}{\scriptsize 2}};
\node at (1.25,0.25) {\textcolor {black!80}{\scriptsize 3}};
\node at (1.75,0.25) {\textcolor {black!80}{\scriptsize 4}};
\node at (0.25,0.75) {\textcolor {black!80}{\scriptsize 5}};
\node at (0.75,0.75) {\textcolor {black!80}{\scriptsize 6}};
\node at (1.25,0.75) {\textcolor {black!80}{\scriptsize 7}};
\node at (1.75,0.75) {\textcolor {black!80}{\scriptsize 8}};
\node at (0.75,1.25) {\textcolor {black!80}{\scriptsize 9}};
\node at (0.25,1.25) {\textcolor {black!80}{\scriptsize 10}};
\node at (2.25,1.25) {\textcolor {black!80}{\scriptsize 10}};
\node at (1.25,1.25) {\textcolor {black!80}{\scriptsize 11}};
\node at (1.75,1.25) {\textcolor {black!80}{\scriptsize 12}};
\node at (0.25,1.75) {\textcolor {black!80}{\scriptsize 13}};
\node at (0.75,1.75) {\textcolor {black!80}{\scriptsize 14}};
\node at (1.25,1.75) {\textcolor {black!80}{\scriptsize 15}};
\node at (1.75,1.75) {\textcolor {black!80}{\scriptsize 16}};

\draw[-,line width=1.25pt](0,0)--(0,2);
\draw[-,line width=1.25pt](0,2)--(2,2);
\draw[-,line width=1.25pt](2,2)--(2,0);
\draw[-,line width=1.25pt](0,0)--(2,0);

\draw[-,line width=1.25pt](0.5,0)--(0.5,2);
\draw[-,line width=1.25pt](1,0)--(1,2);
\draw[-,line width=1.25pt](1.5,0)--(1.5,2);

\draw[-,line width=1.25pt](0,0.5)--(2,0.5);
\draw[-,line width=1.25pt](0,1)--(2,1);
\draw[-,line width=1.25pt](0,1.5)--(2,1.5);

\draw [fill=red,opacity=.4,draw=black!50,line width=1.25pt , dotted] (1.85,0.85) rectangle (2.65,1.65);
\draw [fill=green,opacity=.2,draw=black!50,line width=1.25pt ] (1.35,1.35) rectangle (2.15,2.15);
\draw [fill=green,opacity=.2,draw=black!50,line width=1.25pt ] (1.35,0.85) rectangle (2.15,1.65);
\draw [fill=green,opacity=.2,draw=black!50,line width=1.25pt ] (1.35,0.35) rectangle (2.15,1.15);
\end{tikzpicture}
\end{subfigure}\\[14pt]
\begin{subfigure}{.33\textwidth}
\centering
\begin{tikzpicture}[scale=1.4]
 \draw[-,line width=1.25pt](0,0)--(0,2);
\draw[-,line width=1.25pt](0,2)--(2,2);
\draw[-,line width=1.25pt](2,2)--(2,0);
\draw[-,line width=1.25pt](0,0)--(2,0);

\draw[-,line width=1.25pt](0.5,0)--(0.5,2);
\draw[-,line width=1.25pt](1,0)--(1,2);
\draw[-,line width=1.25pt](1.5,0)--(1.5,2);

\draw[-,line width=1.25pt](0,0.5)--(2,0.5);
\draw[-,line width=1.25pt](0,1)--(2,1);
\draw[-,line width=1.25pt](0,1.5)--(2,1.5);

%

\node at (0.25,0.25) {\textcolor {black!80}{\scriptsize 1}};
\node at (0.75,0.25) {\textcolor {black!80}{\scriptsize 2}};
\node at (1.25,0.25) {\textcolor {black!80}{\scriptsize 3}};
\node at (1.75,0.25) {\textcolor {black!80}{\scriptsize 4}};
\node at (0.25,0.75) {\textcolor {black!80}{\scriptsize 5}};
\node at (0.75,0.75) {\textcolor {black!80}{\scriptsize 6}};
\node at (1.25,0.75) {\textcolor {black!80}{\scriptsize 7}};
\node at (1.75,0.75) {\textcolor {black!80}{\scriptsize 8}};
\node at (0.75,1.25) {\textcolor {black!80}{\scriptsize 9}};
\node at (0.25,1.25) {\textcolor {black!80}{\scriptsize 10}};
\node at (1.25,1.25) {\textcolor {black!80}{\scriptsize 11}};
\node at (1.75,1.25) {\textcolor {black!80}{\scriptsize 12}};
\node at (0.25,1.75) {\textcolor {black!80}{\scriptsize 13}};
\node at (0.75,1.75) {\textcolor {black!80}{\scriptsize 14}};
\node at (1.25,1.75) {\textcolor {black!80}{\scriptsize 15}};
\node at (1.75,1.75) {\textcolor {black!80}{\scriptsize 16}};

\draw [fill=red,opacity=.4,draw=black!80,line width=1.25pt ] (-0.15,-0.15) rectangle (0.65,0.65);
\draw [fill=green,opacity=.2,draw=black!80,line width=1.25pt ] (-0.15,0.35) rectangle (0.65,1.15);
\draw [fill=green,opacity=.2,draw=black!80,line width=1.25pt ] (0.35,0.35) rectangle (1.15,1.15);
\draw [fill=green,opacity=.2,draw=black!80,line width=1.25pt ] (0.35,-0.15) rectangle (1.15,0.65);
\draw [fill=green,opacity=.0,draw=black!00,line width=1.25pt ] (1.64,2.64) rectangle (1.65,2.65);

\end{tikzpicture}
\end{subfigure}%
\begin{subfigure}{.33\textwidth}
\centering
\begin{tikzpicture}[scale=1.4]
\draw[-,line width=1.25pt ](0,0)--(0,2);
\draw[-,line width=1.25pt ](0,2)--(2,2);
\draw[-,line width=1.25pt ](2,2)--(2,0);
\draw[-,line width=1.25pt](0,0)--(2,0);

\draw[-,line width=1.25pt ](-0.15,0.65)--(-0.15,-0.15);
\draw[-,line width=1.25pt ](-0.15,-0.15)--(0.65,-0.15);
\draw[-,line width=1.25pt ](0.65,-0.15)--(0.65,0.65);
\draw[-,line width=1.25pt ](0.65,0.65)--(-0.15,0.65);

\draw[dotted,line width=1.25pt ](1.85,0.65)--(1.85,-0.15);
\draw[dotted,line width=1.25pt ](1.85,-0.15)--(2.65,-0.15);
\draw[dotted,line width=1.25pt ](2.65,-0.15)--(2.65,0.65);
\draw[dotted,line width=1.25pt ](2.65,0.65)--(1.85,0.65);

\draw[dotted,line width=1.25pt ](-0.15,2.65)--(-0.15,1.85);
\draw[dotted ,line width=1.25pt](-0.15,1.85)--(0.65,1.85);
\draw[dotted,line width=1.25pt ](0.65,1.85)--(0.65,2.65);
\draw[dotted ,line width=1.25pt](0.65,2.65)--(-0.15,2.65);

\draw[dotted ,line width=1.25pt](1.85,2.65)--(1.85,1.85);
\draw[dotted ,line width=1.25pt](1.85,1.85)--(2.65,1.85);
\draw[dotted ,line width=1.25pt](2.65,1.85)--(2.65,2.65);
\draw[dotted ,line width=1.25pt](2.65,2.65)--(1.85,2.65);

\draw[->,line width=1.25pt,black!0](0.39,0.65)--node[black,sloped, anchor=center,above]{\scriptsize $T_{1,2}(\omega_i)$}(0.39,1.85);
\draw[->,line width=1.25pt,black!0 ](0.65,0.31)--node[black,above]{\scriptsize $T_{1,1}(\omega_i)$}(1.85,0.31);
\draw[->,line width=1.25pt,black!0 ](0.65,0.61)--node[black,sloped, anchor=center,above]{\scriptsize $T_{1,3}(\omega_i)$}(1.85,1.81);

\draw[->,line width=1.25pt](0.35,0.65)--(0.35,1.85);
\draw[-> ,line width=1.25pt](0.65,0.35)--(1.85,0.35);
\draw[-> ,line width=1.25pt](0.65,0.65)--(1.85,1.85);

\node at (0.25,0.25) {\textcolor {black!80}{\scriptsize 1}};
\node at (2.25,0.25) {\textcolor {black!80}{\scriptsize 1}};
\node at (2.25,2.25) {\textcolor {black!80}{\scriptsize 1}};
\node at (0.25,2.25) {\textcolor {black!80}{\scriptsize 1}};
\end{tikzpicture}
\end{subfigure}%
\hfill
\begin{subfigure}{.33\textwidth}
\centering
\begin{tikzpicture}[scale=1.3]
\draw[-,line width=1.25pt](0,0)--(0,2);
\draw[-,line width=1.25pt](0,2)--(2,2);
\draw[-,line width=1.25pt](2,2)--(2,0);
\draw[-,line width=1.25pt](0,0)--(2,0);

\draw[-,line width=1.25pt](0.5,0)--(0.5,2);
\draw[-,line width=1.25pt](1,0)--(1,2);
\draw[-,line width=1.25pt](1.5,0)--(1.5,2);

\draw[-,line width=1.25pt](0,0.5)--(2,0.5);
\draw[-,line width=1.25pt](0,1)--(2,1);
\draw[-,line width=1.25pt](0,1.5)--(2,1.5);

\node at (0.25,0.25) {\textcolor {black!80}{\scriptsize 1}};
\node at (2.25,0.25) {\textcolor {black!80}{\scriptsize 1}};
\node at (2.25,2.25) {\textcolor {black!80}{\scriptsize 1}};
\node at (0.25,2.25) {\textcolor {black!80}{\scriptsize 1}};
\node at (0.75,0.25) {\textcolor {black!80}{\scriptsize 2}};
\node at (1.25,0.25) {\textcolor {black!80}{\scriptsize 3}};
\node at (1.75,0.25) {\textcolor {black!80}{\scriptsize 4}};
\node at (0.25,0.75) {\textcolor {black!80}{\scriptsize 5}};
\node at (0.75,0.75) {\textcolor {black!80}{\scriptsize 6}};
\node at (1.25,0.75) {\textcolor {black!80}{\scriptsize 7}};
\node at (1.75,0.75) {\textcolor {black!80}{\scriptsize 8}};
\node at (0.75,1.25) {\textcolor {black!80}{\scriptsize 9}};
\node at (0.25,1.25) {\textcolor {black!80}{\scriptsize 10}};
\node at (1.25,1.25) {\textcolor {black!80}{\scriptsize 11}};
\node at (1.75,1.25) {\textcolor {black!80}{\scriptsize 12}};
\node at (0.25,1.75) {\textcolor {black!80}{\scriptsize 13}};
\node at (0.75,1.75) {\textcolor {black!80}{\scriptsize 14}};
\node at (1.25,1.75) {\textcolor {black!80}{\scriptsize 15}};
\node at (1.75,1.75) {\textcolor {black!80}{\scriptsize 16}};

\draw [fill=red,opacity=.4,draw=black!80,line width=1.25pt, dotted ] (1.85,-0.15) rectangle (2.65,0.65);
\draw [fill=red,opacity=.4,draw=black!80,line width=1.25pt,dotted ] (-0.15,1.85) rectangle (0.65,2.65);
\draw [fill=red,opacity=.4,draw=black!80,line width=1.25pt,dotted ] (1.85,1.85) rectangle (2.65,2.65);

\draw [fill=green,opacity=.2,draw=black!80,line width=1.25pt ] (-0.15,1.35) rectangle (0.65,2.15);
\draw [fill=green,opacity=.2,draw=black!80,line width=1.25pt ] (0.35,1.35) rectangle (1.15,2.15);
\draw [fill=green,opacity=.2,draw=black!80,line width=1.25pt ] (1.35,1.35) rectangle (2.15,2.15);
\draw [fill=green,opacity=.2,draw=black!80,line width=1.25pt ] (1.35,0.35) rectangle (2.15,1.15);
\draw [fill=green,opacity=.2,draw=black!80,line width=1.25pt ] (1.35,-0.15) rectangle (2.15,0.65);
\end{tikzpicture}
\end{subfigure}
\caption{Schematic of the implementation of the periodicity of a PU function $\varphi_i$ \eqref{equation:shepardFunction} associated with a patch $\omega_i$ (red) by generalization of 
the notion of neighboring or overlapping patches $\omega_i \cap \omega_j\not=\emptyset$ (green) at the boundary $\partial\Omega$ via periodic 
copies of patches $\omega_i \cap \partial \Omega \not=\emptyset$ (red), see~\cite{Klaar:2016} for details.}
\label{figure:periodic_neighbors}
\end{figure}

The main difference between our PUM approach~\cite{MASchweitzer:2003,MASchweitzer:Habilitation} and most other generalized or extended FEM techniques, 
compare~\cite{Fries:Belytschko:2010,Schweitzer:2012}, is that we employ a so-called flat-top PU instead of classical linear Lagrange FE functions, compare~\fref{figure:comparePU}. For the 
construction of such a flat-top PU let us first define a cover $C_{\Omega} \coloneqq \{\omega_i\}$ of the domain $\Omega$ with the help of a uniform regular mesh of mesh-width $2h$ by an isotropic 
scaling of the mesh-cells
\begin{displaymath}
{\mathcal C}_i = \prod_{l=1}^d (o_i^l-h, o_i^l+h), 
\end{displaymath}
i.e., we define the patches $\omega_i$ as 
\begin{equation}\label{equation:coverPatch}
\omega_i \coloneqq \prod_{l=1}^d (o_i^l-\alpha h, o_i^l+\alpha h),\quad \text{with } \alpha > 1,
\end{equation}
see Figure~\ref{figure:uniformCover}.
To obtain a PU on a cover $C_\Omega$ with $N\coloneqq\operatorname{card}(C_\Omega)$ we define a weight function
$W_{i}:\Omega \rightarrow \mathbb{R}$ with $\operatorname{supp}(W_{i})=\omega_{i}$ for each cover patch
$\omega_{i}$ by
\begin{equation}\label{equation:weightFunction}
W_{i}(x) = \left\{ \begin{array}{l@{\hspace{1.0em}}l} \mathcal{W} \circ T_{i} (x), & x \in \omega_{i} \\ 0, & \text{otherwise} \end{array} \right.
\end{equation}
with the affine transforms $T_{i}: \overline{\omega}_{i} \rightarrow [-1,1]^d$ and $\mathcal{W}:[-1,1]^d\rightarrow\mathbb{R}$ any non-negative compactly supported function, such as a B-spline. By normalizing these weight functions we obtain the functions
\begin{equation}
\label{equation:shepardFunction}
\varphi_{i}(x) \coloneqq \frac{W_{i}(x)}{\sum_{l \in C_i} W_{l}(x)},
\end{equation}
where we define the local neighborhood $C_i\coloneqq\{l : \omega_l \cap \omega_i \not=\emptyset\}$ of a patch $\omega_i$ in a slightly more general way to account for the required periodicity of the 
basis functions, see Figures \ref{figure:periodic_neighbors_1D} and \ref{figure:periodic_neighbors}. To impose Bloch-periodic boundary conditions, we multiply the entries of the resulting system 
matrices in matrix blocks that correspond to Bloch-boundary patches with the appropriate Bloch phase factor. This results in a periodic PU and Bloch-periodic basis, analogous to the procedure in~\cite{Sukumar:Pask:2009}. 
Note that the PU~\eqref{equation:shepardFunction} is non-negative since the employed weight functions are non-negative
and that the $\varphi_i$ satisfy the flat-top property for any $\alpha \in (1,2)$, see~\cite{MGriebel:MASchweitzer:2006,MASchweitzer:2007a}. Due to this construction we can easily control the overlap of the
patches and thereby the size of the flat-top region by the parameter $\alpha \in (1,2)$. Note that for $\alpha=1$ our PUM degenerates to a discontinuous Galerkin approach whereas with the choice of $\alpha=2$ and linear B-spline weights 
in~\eqref{equation:weightFunction} our PUM yields the PUFEM.

\subsection{Orbital enrichment and stability}
\label{section:orbital_enrichment_and_stability}
The approximation power of an enriched PUM is mostly obtained by the choice of high quality enrichment space $V^\mathcal{E}$ (enrichment functions $\psi_i^t$) on the patches $\omega_i$ for the 
problem at hand. For instance, the use of generalized harmonic polynomials or local spaces based on planewaves have been employed successfully for smooth problems in~\cite{BaMe:96, Strouboulis20064711, Strouboulis2008364}, whereas for crack propagation problems the use of discontinuous and singular enrichment functions is appropriate, 
see e.g.~\cite{DuOd:95, TBelytschko:YYLu:LGu:1995,Schweitzer:2012}. In the context of the Schr\"odinger and Poisson equations of Kohn--Sham density functional theory, enrichment functions 
$\psi_i^t$ constructed 
from isolated-atom solutions were employed in~\cite{Pask:Sukumar:Guney:Hu:2011,Pask:Sukumar:Mousavi:2012,Pask:Sukumar:2017}, as in previous work~\cite{Sukumar:Pask:2009} in the context of model densities and potentials. 
Our general flat-top PUM allows for the use of arbitrary problem-dependent enrichment functions and thus we anticipate 
that our PUM has essentially the same convergence properties as for example the
PUFEM of~\cite{Sukumar:Pask:2009,Pask:Sukumar:Mousavi:2012,Pask:Sukumar:2017} when we employ the same enrichment functions.

The fundamental difference of our approach from the PUFEM of~\cite{Sukumar:Pask:2009,Pask:Sukumar:Mousavi:2012,Pask:Sukumar:2017} is that we employ a flat-top 
PU~\cite{MASchweitzer:2008d,MASchweitzer:2002,MASchweitzer:Habilitation} to overcome the two major challenges encountered in the PUFEM: ill-conditioning of the overlap matrix and the need 
for the solution of a generalized eigenproblem.

Let us first consider the ill-conditioning of the overlap matrix, i.e., the $L^2$-stability of~\eqref{equation:pumSpaceDefinition}, which can be encountered in all enriched 
approximations~\cite{AHuerta:TBelytschko:TFernandezMendez:TRabczuk:2004,NMoes:JDolbow:TBelytschko:1999,Fries:Belytschko:2010,Schweitzer:2012,Babuska:Banerjee:2011}. 
For the smooth space $V^\mathcal{P}_i$ in~\eqref{equation:localApproximationSpaceDefinition} we employ a local basis $\pi_i^s$ on $\omega_i$,
i.e., $\pi_i^s = p_s \circ T_i$ and $\{p_s\}$ denotes a stable basis on $[-1,1]^d$, for instance Legendre-polynomials. The enrichment functions
$\psi_i^t$ however are often given as global functions $\eta^t$ on the computational domain $\Omega$ since they are
designed to capture special behavior of the global solution at a particular location. Therefore, the restrictions $\psi_i^t
\coloneqq \eta^t|_{\omega_i}$ of the enrichment functions $\eta^t$ to a particular patch $\omega_i$ may be
ill-conditioned or even linearly dependent on $\omega_i$, even if the enrichment functions $\eta^t$ are
well-conditioned on the global domain. Furthermore, the coupling between the spaces $V^\mathcal{P}_i$ and
$V^\mathcal{E}_i$ on the patch $\omega_i$ must be considered. The set of functions $\{\pi_i^s,
\psi_i^t\}$ will degenerate from being a basis of $V_i$ to yielding a generating system if the restricted enrichment
functions $\psi_i^t=\eta^t|_{\omega_i}$ can be well-approximated by polynomials $\pi_i^s$ on the patch
$\omega_i$. Moreover, in the general case when~\eqref{equation:pumSpaceDefinition} employs an arbitrary non-flat-top PU, e.g., a PU built from FE, we also need to consider the interactions of the overlapping 
local approximation spaces $V_i$ and $V_j$ \eqref{equation:localApproximationSpaceDefinition} on neighboring patches $\omega_i \cap \omega_j \not= \emptyset$ which essentially introduces global 
constraints and therefore renders the use of an arbitrary non-flat-top PU computationally infeasible. If we, however, restrict ourselves to the use of a flat-top PU, all issues that may result in 
ill-conditioning can be efficiently resolved solely on the local patches $\omega_i$, see~\cite{MASchweitzer:2008d} for further details. 

To attain a stable basis of $V^{\mathrm{PU}}$ for arbitrary local approximation spaces $V_i$ with a flat-top PU $\{\varphi_i\}$, we essentially need to be able to identify a local enrichment 
space $V^\mathcal{D}_i$ such that we can rewrite~\eqref{equation:localApproximationSpaceDefinition} as 
\begin{displaymath}
V_i = V^\mathcal{P}_i + V^\mathcal{E}_i = V^\mathcal{P}_i \oplus V^\mathcal{D}_i.
\end{displaymath}
To this end, we choose an inner product on the patch $\omega_i$ and construct $V^\mathcal{D}_i$ such that $V^\mathcal{P}_i \perp V^\mathcal{D}_i$ holds. The respective stability transformation
\begin{equation}
\label{equation:stabilityTransform}
P: V^{\mathrm{PU}} = \sum_{i=1}^N \varphi_i (V^\mathcal{P}_i + V^\mathcal{E}_i) \rightarrow \sum_{i=1}^N \varphi_i (V^\mathcal{P}_i \oplus V^\mathcal{D}_i)
\end{equation}
can in fact be computed efficiently on-the-fly by partial orthogonalization with respect to the chosen inner product and only involves local operations on the patches~\cite{MASchweitzer:2008d}.
The particular stability transformation on a patch $\omega_i$ depends on the choice of the inner product and works in four steps starting from a local matrix 
\begin{displaymath}
M_{i} = \left(\begin{array}{cc} M_{i;\mathcal{P},\mathcal{P}} & M_{i;\mathcal{P},\mathcal{E}} \\
    M_{i;\mathcal{E},\mathcal{P}} & M_{i;\mathcal{E},\mathcal{E}} \end{array}\right)
\end{displaymath}
arising from Gram matrix of all basis functions $\pi_i^s$ of $V^\mathcal{P}_i$ and $\psi_i^t$ of $V^\mathcal{E}_i$ in the chosen inner product. 
First, all basis functions of $V^\mathcal{P}_i$ and $V^\mathcal{E}_i$ are scaled appropriately with respect to the employed inner product, so that the induced norm of all functions is identical. 
Then, we compute eigenvalue decompositions of $M_{i;\mathcal{P},\mathcal{P}}$ and $M_{i;\mathcal{E},\mathcal{E}}$ individually and use thresholding of small eigenvalues to eliminate any instabilities within $V^\mathcal{P}_i$ and $V^\mathcal{E}_i$, so that we can transform $M_{i}$ to attain the form
\begin{displaymath}
M^{(1)}_{i} = \left(\begin{array}{cc} \mathbb{I}_{\mathcal{P}} & M^{(1)}_{i;\mathcal{P},\mathcal{E}} \\
    M^{(1)}_{i;\mathcal{E},\mathcal{P}} & \mathbb{I}_{\mathcal{E}} \end{array}\right).
\end{displaymath}
Then, the overlap of the two spaces $V_i^{\mathcal{P}}$ 
and $V_i^{\mathcal{E}}$ is removed by partial orthogonalization (Schur complement) which yields
\begin{displaymath}
M^{(2)}_{i} = \left(\begin{array}{cc} \mathbb{I}_{\mathcal{P}} & 0 \\
    0 & M^{(2)}_{i;\mathcal{D},\mathcal{D}}    \end{array}\right).
\end{displaymath}
Thus, with the help of an eigenvalue decomposition of $M^{(2)}_{i;\mathcal{D},\mathcal{D}}$ we obtain an orthonormal basis of $V_i$ with respect to the employed inner product.
And so the problem of ill-conditioning which is encountered in PUFEM can be resolved in a flat-top PUM while fully maintaining the improved approximation quality due to enrichment.

\subsection{Variational mass lumping}
\label{section:variational_mass_lumping}
The second drawback of the PUFEM identified in~\cite{Pask:Sukumar:2017} is related to the fact that the Galerkin discretization of~\eqref{equation:schroedinger_strong} yields a Hermitian 
generalized eigenproblem~\eqref{equation:generalized_eigenvalue_problem} where $H$ and $S$ are in general sparse but not diagonal matrices, which renders the solution of~\eqref{equation:generalized_eigenvalue_problem} more challenging since the computation of $S^{-1}$ is implicitly required, i.e., one effectively needs to solve the standard eigenproblem
\begin{displaymath}
S^{-1}H \tilde{u} = \varepsilon \tilde{u}.
\end{displaymath}
In the context of classical (linear) FEM one therefore often replaces the overlap (or consistent mass) matrix $S$ by an approximation $\bar{S}$ whose inverse $\bar{S}^{-1}$ can be 
computed much more efficiently, e.g., if  $\bar{S}$ is diagonal, by so-called mass-lumping~\cite{Hu:Huang:Shen:2004,StrangFix:1973,Tong:Pian:Bucciarblli:1971}. Various constructions 
for $\bar{S}$ exist in the FEM context~\cite{hughes2000finite}, yet
these approaches are in general not directly applicable to a PUM or any enriched approximation scheme due to the
non-polynomial and non-interpolatory character of the employed basis functions.

Fortunately, there is a natural and rather simple approach to the construction of an appropriate approximation of the
overlap matrix for the PUM in general~\cite{Schweitzer:2013*2}. It is applicable to any non-negative PU
$\{\varphi_i\}$ and arbitrary local approximation spaces~\eqref{equation:localApproximationSpaceDefinition}
(e.g., higher order polynomials, discontinuous, and singular enrichments) and utilizes only the special product structure of the PUM basis functions $\varphi_i \vartheta_i^n$.

To introduce this variational lumping scheme for our PUM, recall that the consistent overlap matrix is given by 
\begin{equation}
\label{equation:consistent-mass-matrix-pum}
S = (S_{(i,n),(j,m)}), \quad S_{(i,n),(j,m)} = \langle \varphi_j\vartheta_j^m, \varphi_i\vartheta_i^n\rangle_{L^2(\Omega,\mathbb{C})},
\end{equation}
where our flat-top PU functions $\varphi_i$ and $\varphi_j$ are non-negative. Thus, the entries of the overlap matrix are given by the inner products in $L^2(\Omega,\mathbb{C})$ of our 
PUM basis functions. In~\cite{Schweitzer:2013*2} it was shown that the global $L^2(\Omega,\mathbb{C})$ inner product in~\eqref{equation:consistent-mass-matrix-pum} can be replaced by local 
weighted $L^2(\omega_i,\mathbb{C}; \varphi_i)$ 
inner products without diminishing the global convergence behavior, since there holds the equivalence 
\begin{displaymath}
\langle f, \varphi_i \vartheta_i^n\rangle_{L^2(\Omega,\mathbb{C})} = \int_\Omega f \varphi_i \vartheta_i^n ~dx =
\int_{\Omega \cap \omega_i} f \varphi_i \vartheta_i^n ~dx =: \langle f, \vartheta_i^n\rangle_{L^2(\Omega \cap \omega_i, \mathbb{C}; \varphi_i)}
\end{displaymath}
for an arbitrary function $f \in L^2(\Omega, \mathbb{C})$. 
With the help of these local inner products, we therefore obtain an approximate overlap matrix 
\begin{equation}
\label{equation:lumped-mass-matrix-pum}
\bar{S} =  (\bar{S}_{(i,n),(j,m)}), \quad \bar{S}_{(i,n),(j,m)} = \left\{\begin{array}{cl} 0 & i \not= j\\ \langle
  \vartheta_i^m, \vartheta_i^n\rangle_{L^2(\Omega \cap \omega_i, \mathbb{C}; \varphi_i)} & i=j
\end{array}\right.
\end{equation}
which is block-diagonal and symmetric positive definite for any choice of the local approximation spaces $V_i$, see~\cite{Schweitzer:2013*2} for further details. Therefore, an approximate 
solution of~\eqref{equation:generalized_eigenvalue_problem} can be obtained very efficiently via
\begin{equation}
\label{equation:generalized_eigenvalue_problem_lumped}
H \tilde{u} = \varepsilon \bar{S} \tilde{u}.
\end{equation}
In fact, choosing the local weighted $L^2(\omega_i,\mathbb{C}; \varphi_i)$ inner products also in the stability transformation~\eqref{equation:stabilityTransform} 
yields $\bar{S}=\mathbb{I}$, so that~\eqref{equation:generalized_eigenvalue_problem_lumped} becomes $H \tilde{u} = \varepsilon \tilde{u}$ and 
the expensive solution of the generalized eigenproblem encountered in the PUFEM can be completely circumvented in our flat-top PUM. 

\section{Numerical results}
\label{section:numerical_results}
In this section we present some numerical results obtained with our flat-top PUM. 
We consider two benchmark problems defined on three-dimensional unit cells~\cite{Sukumar:Pask:2009} to validate our implementation and assess the accuracy and efficiency in terms 
of the degrees of freedom (DOFs) of the flat-top PUM with respect to the Schr\"odinger eigenproblem with periodic and Bloch-periodic boundary conditions. The two main objects of interest of our study, however, are the 
conditioning of the consistent and lumped overlap matrices (before and after the stabilization~\eqref{equation:stabilityTransform}) and the effect our lumping scheme~\eqref{equation:lumped-mass-matrix-pum} has on the quality of the approximation. Since absolute runtime is not the focus of this study, we employed a 
simple tensor-product $6\times6\times6$ Gau{\ss} integration 
rule on subdivided cover cells~\cite{MASchweitzer:2003,MGriebel:MASchweitzer:2001a}.

\subsection{Three-dimensional quantum harmonic oscillator}
\label{subsection:HO}
As in~\cite{Sukumar:Pask:2009}, the first benchmark problem we consider is the Schr\"odinger equation~\eqref{equation:schroedinger_strong} with harmonic potential
\begin{equation}
\label{equation:harmonic_potential}
V\left(|\vm{x}-\vm{\tau}|\right) = \frac{|\vm{x}-\vm{\tau}|^2}{2},
\end{equation}
under periodic boundary conditions ($\vm{k}=0$). Note that this potential itself is not periodic. As our unit cell $\Omega$, 
we take a cuboid cell with primitive lattice vectors $\vm{a}_1 \coloneqq a(1, 0, 0)$, $\vm{a}_2 \coloneqq a(0, 1.1, 0)$ and 
$\vm{a}_3 \coloneqq a(0, 0, 1.2)$. As lattice parameter we choose $a=5~\text{au}$ and the potential 
center is given by $\vm{\tau} \coloneqq \frac{\vm{a}_1+\vm{a}_2+\vm{a}_3}{2}$. 

On each patch, the polynomial approximation space $V^\mathcal{P}_i$ contains Legendre polynomials up to total degree $p = 1,2,3$, while the enrichment space $V^\mathcal{E}_i$ consists of periodic 
lattice sums of the infinite box eigenfunctions $\psi^{\prime}_{nlm}(\vm{x}) = R_{nl}(r)Y_{lm}(\theta, \phi)$, 
\begin{equation}
 \psi_{nlm} = \sum_{\vm{R}} \psi^{\prime}_{nlm}(|\vm{x}-\vm{\tau}-\vm{R}|),
\end{equation}
over lattice translation vectors $\vm{R} = i_1 \vm{a}_1+i_2 \vm{a}_2+i_3 \vm{a}_3$, to enrich each of the ten lowest states. Here, $Y_{lm}$ are the analytically given 
spherical harmonics and $R_{nl}$ is the 
radial part of the eigenfunction, which is obtained by solving the radial 
Schr\"odinger equation~\cite{Sukumar:Pask:2009}. Thus, the radial components are given numerically, i.e., at discrete points.
To ensure a compact support of $R_{nl}$ we multiply its discrete values by a $\mathcal{C}^3$ cut-off function~\cite{Sukumar:Pask:2009}
\begin{equation}
\label{equation:cut_off_function}
 h(r, r_0) \coloneqq \left\{
 \begin{aligned}
  &1 + \frac{20r^7}{r_0^7} -\frac{70r^6}{r_0^6} + \frac{84r^5}{r_0^5} -\frac{35r^4}{r_0^4} , & r\leq r_0 \\
  &0,        & r > r_0 
 \end{aligned}
 \right. ,
\end{equation}
where we choose $r_0 = 6~\text{au}$. The resulting product is then interpolated by a quintic spline.

\begin{figure}
 \begin{center}
     \begin{tikzpicture}[scale=0.65]
  \draw [very thin ] (0.0,0.0) rectangle (6.0,5.0);
   \foreach \x in {6/7, 12/7, 18/7, 24/7, 30/7, 36/7, 6} {
    \draw[-,very thin](\x,0)--(\x,5);}
   \foreach \x in {5/7, 10/7, 15/7, 20/7, 25/7, 30/7, 5} {
    \draw[-,very thin](0, \x)--(6, \x);}
   \filldraw (3, 2.5) circle (1.5pt) node[align=right,  right] {$\vm{\tau}$};
 \end{tikzpicture}
   \begin{tikzpicture}[scale=0.65]
    \draw [very thin ] (0.0,0.0) rectangle (6.0,5.0);
   \foreach \x in {6/7, 12/7, 18/7, 24/7, 30/7, 36/7, 6} {
    \draw[-,very thin](\x,0)--(\x,5);}
   \foreach \x in {5/7, 10/7, 15/7, 20/7, 25/7, 30/7, 5} {
    \draw[-,very thin](0, \x)--(6, \x);}
   \filldraw (3, 2.5) circle (1.5pt) node[align=right,  right] {$\vm{\tau}$};
   \draw[color=red] (3, 2.5) circle [radius=1.0];
    \draw [fill=red,opacity=.2,draw=red!20,very thin] (12/7,15/7) rectangle (30/7, 20/7);
    \draw [fill=red,opacity=.2,draw=red!20,very thin] (18/7,10/7) rectangle (24/7, 15/7);
    \draw [fill=red,opacity=.2,draw=red!20,very thin] (18/7,20/7) rectangle (24/7, 25/7);
 \end{tikzpicture}
    \begin{tikzpicture}[scale=0.65]
  \draw [very thin ] (0.0,0.0) rectangle (6.0,5.0);
   \foreach \x in {6/7, 12/7, 18/7, 24/7, 30/7, 36/7, 6} {
    \draw[-,very thin](\x,0)--(\x,5);}
   \foreach \x in {5/7, 10/7, 15/7, 20/7, 25/7, 30/7, 5} {
    \draw[-,very thin](0, \x)--(6, \x);}
   \filldraw (3, 2.5) circle (1.5pt) node[align=right,  right] {$\vm{\tau}$};
   \draw[color=red] (3, 2.5) circle [radius=2.5];
    \draw [fill=red,opacity=.2,draw=red!20,very thin] (6/7,5/7) rectangle (36/7, 30/7);
    \draw [fill=red,opacity=.2,draw=red!20,very thin] (12/7,0) rectangle (30/7, 5/7);
    \draw [fill=red,opacity=.2,draw=red!20,very thin] (12/7,30/7) rectangle (30/7, 5);
 \end{tikzpicture}
 
\end{center}
\caption{Sketch of refinement by increasing the enrichment radius $r_e$ in two dimensions. We only enrich a patch if its center point lies within the ball of radius $r_e$ centered at the potential center $\vm{\tau}$. 
Here, this ball and enriched patches are colored in red. Depicted are three different values for $r_e$: $0.0~\text{au}$ (left), $1.0~\text{au}$ (center) and $2.5~\text{au}$ (right) on a $7\times7$ cover.}
\label{figure:enrichment_radius}
\end{figure}
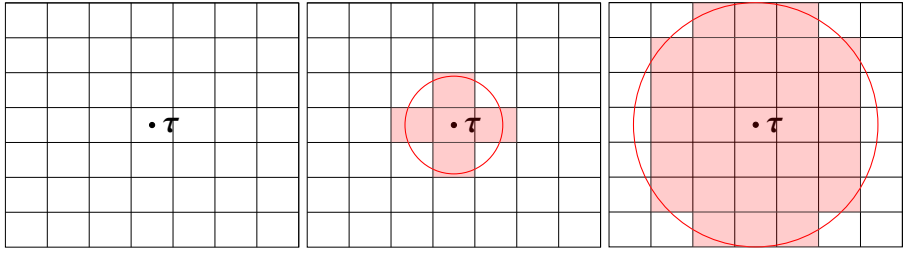
\begin{figure}[tbp]
 \centering
 \tikzset{every mark/.append style={scale=1.25}}
 \begin{tikzpicture}[scale=1.0]
  \begin{loglogaxis}[
   xlabel={degrees of freedom},
   ylabel={absolute error (Ha)},
   ymin=8e-08,
   grid=major,
   legend style={font=\tiny},legend pos=south west,
   legend entries={$p=1$, $p=2$, $p=3$, $(p=1) + e(\uparrow r_e)$, $(p=2) + e(\uparrow r_e)$, $(p=3) + e(\uparrow r_e)$},
   ]
   \addplot[teal,mark=triangle,line width=0.8pt] table [x=dof, y=e_0] {Data/gswithoutenrich/Abs_errors_Harmonic_Oscillator_Consistent_poldeg1.dat};
   \addplot[lime,mark=square,line width=0.8pt] table [x=dof, y=e_0] {Data/gswithoutenrich/Abs_errors_Harmonic_Oscillator_Consistent_poldeg2.dat};
   \addplot[magenta,mark=o,line width=0.8pt] table [x=dof, y=e_0] {Data/gswithoutenrich/Abs_errors_Harmonic_Oscillator_Consistent_poldeg3.dat};
   \addplot[blue,mark=triangle*,line width=0.8pt] table [x=dof, y=e_0] {Data/groundstateincrradiusoddp/Abs_Errors_Harmonic_Oscillator_Consistent_enriched_poldeg1_incr_radius_l3_numenr1_smalleps.dat};
   \addplot[green,mark=square*,line width=0.8pt] table [x=dof, y=e_0] {Data/groundstateincrradiusoddp/Abs_Errors_Harmonic_Oscillator_Consistent_enriched_poldeg2_incr_radius_l3_numenr1_smalleps.dat};
   \addplot[red,mark=*,line width=0.8pt] table [x=dof, y=e_0] {Data/groundstateincrradiusoddp/Abs_Errors_Harmonic_Oscillator_Consistent_enriched_poldeg3_incr_radius_l3_numenr1_smalleps.dat};
   \addplot[blue,mark=triangle*,line width=0.8pt] table [x=dof, y=e_0] {Data/groundstateincrradiusoddp/Abs_Errors_Harmonic_Oscillator_Consistent_enriched_poldeg1_incr_radius_l2_numenr1_smalleps.dat};
   \addplot[green,mark=square*,line width=0.8pt] table [x=dof, y=e_0] {Data/groundstateincrradiusoddp/Abs_Errors_Harmonic_Oscillator_Consistent_enriched_poldeg2_incr_radius_l2_numenr1_smalleps.dat};
   \addplot[red,mark=*,line width=0.8pt] table [x=dof, y=e_0] {Data/groundstateincrradiusoddp/Abs_Errors_Harmonic_Oscillator_Consistent_enriched_poldeg3_incr_radius_l2_numenr1_smalleps.dat};
   \addplot[black,sharp plot,update limits=false,line width=1pt] coordinates {(10,1.00E-03) (10000000,1.00E-03)};
  \end{loglogaxis}
 \end{tikzpicture}\\[15pt]
 \begin{tikzpicture}[scale=1.0]
  \begin{loglogaxis}[
   xlabel={degrees of freedom},
   ylabel={absolute error (Ha)},
   ymin=5e-08,
   grid=major,
   legend style={font=\tiny},legend pos=south west,
   legend entries={$p=1$, $p=2$, $p=3$, $(p=1) + e(\uparrow r_e)$, $(p=2) + e(\uparrow r_e)$, $(p=3) + e(\uparrow r_e)$},
   ]
   \addplot[teal,mark=triangle,line width=0.8pt] table [x=dof, y=e_0] {Data/gswithoutenrich/Abs_errors_Harmonic_Oscillator_Consistent_poldeg1.dat};
   \addplot[lime,mark=square,line width=0.8pt] table [x=dof, y=e_0] {Data/gswithoutenrich/Abs_errors_Harmonic_Oscillator_Consistent_poldeg2.dat};
   \addplot[magenta,mark=o,line width=0.8pt] table [x=dof, y=e_0] {Data/gswithoutenrich/Abs_errors_Harmonic_Oscillator_Consistent_poldeg3.dat};
   \addplot[blue,mark=triangle*,line width=0.8pt] table [x=dof, y=e_0] {Data/groundstateincrradius/Abs_Errors_Harmonic_Oscillator_Consistent_enriched_poldeg1_incr_radius_l3_numenr1_smalleps.dat};
   \addplot[green,mark=square*,line width=0.8pt] table [x=dof, y=e_0] {Data/groundstateincrradius/Abs_Errors_Harmonic_Oscillator_Consistent_enriched_poldeg2_incr_radius_l3_numenr1_smalleps.dat};
   \addplot[red,mark=*,line width=0.8pt] table [x=dof, y=e_0] {Data/groundstateincrradius/Abs_Errors_Harmonic_Oscillator_Consistent_enriched_poldeg3_incr_radius_l3_numenr1_smalleps.dat};
   \addplot[blue,mark=triangle*,line width=0.8pt] table [x=dof, y=e_0] {Data/groundstateincrradius/Abs_Errors_Harmonic_Oscillator_Consistent_enriched_poldeg1_incr_radius_l2_numenr1_smalleps.dat};
   \addplot[green,mark=square*,line width=0.8pt] table [x=dof, y=e_0] {Data/groundstateincrradius/Abs_Errors_Harmonic_Oscillator_Consistent_enriched_poldeg2_incr_radius_l2_numenr1_smalleps.dat};
   \addplot[red,mark=*,line width=0.8pt] table [x=dof, y=e_0] {Data/groundstateincrradius/Abs_Errors_Harmonic_Oscillator_Consistent_enriched_poldeg3_incr_radius_l2_numenr1_smalleps.dat};
   \addplot[black,sharp plot,update limits=false,line width=1pt] coordinates {(10,1.00E-03) (10000000,1.00E-03)};
  \end{loglogaxis}
 \end{tikzpicture}
 \caption{Convergence history of the lowest eigenvalue $\lambda_1$ for the harmonic oscillator potential~\eqref{equation:harmonic_potential} attained for different refinement schemes. We consider a 
 purely polynomial approximation ($p=1,2,3$) on a sequence of uniformly refined covers, and a refinement by increasing the enrichment radius (compare~\fref{figure:enrichment_radius}) with a single 
 enrichment function on a fixed uniform cover (top: $3\times3\times3$, $7\times7\times7$; bottom: $4\times4\times4$, $8\times8\times8$) that is labeled by
$p=1,2,3 + e\uparrow r_e$. All results were obtained with 
 the consistent overlap matrix.}
 \label{figure:HO_consistent_radii}
\end{figure}

For our numerical experiments, we measure the absolute error of the computed eigenvalues with respect to a cubic finite element reference solution, computed on a $64\times64\times64$ mesh 
which is accurate to $7$ digits (and yields $\lambda_1^{\textrm{ref}}=1.4917524~\text{Ha}$ and $\sum_{i=1}^{10} \lambda_i^{\textrm{ref}} = 29.7715084~\text{Ha}$). 

First, we assess the accuracy of the consistent method. To this end, we first apply our flat-top PUM with linear, quadratic and cubic polynomial spaces on uniformly refined covers without any orbital 
enrichment. As stretch factor~\eqref{equation:coverPatch} we choose $\alpha=1.1$. Here, we anticipate that the flat-top PUM converges with rates that are comparable to classical finite element methods 
of the respective order. Then, we use a fixed uniform cover and use 
enriched local approximation spaces, where we increase the enrichment support radius from $0$ to $4~\text{au}$ (see~\fref{figure:enrichment_radius}). In the first case we do not use any enrichments at all, 
in the latter we enrich each patch whose center lies within the support radius of the enrichment center $\vm{\tau}$. The results are shown in~\fref{figure:HO_consistent_radii}. From the depicted 
plots we can see that the convergence behavior of the non-enriched method is just as expected, while it is also evident that using enrichment functions is very beneficial in terms of accuracy: 
significantly less than 500 DOFs (e.g., 320 DOFs with linear polynomials on a $4\times4\times4$ cover and 297 DOFs with quadratic polynomials on a $3\times3\times3$ cover) yield an error of less than 
$10^{-3}$ Ha, which is a typical requirement in quantum mechanical calculations. Apart from reducing DOFs, enrichment also reduces the overall wall clock time of the method: in our experiments, 
the computation (including setup, assembling of matrices and solving the EVP) on a $3\times3\times3$ cover with quadratic polynomials, a single enrichment function for the lowest eigenvalue and an 
enrichment support radius of $4~\text{au}$ took $7.03$ seconds on $8$ cores, while the non-enriched computation with quadratic polynomials on a $16\times16\times16$ cover took $192.67$ seconds on $8$ cores. 
Both methods yield about the same absolute error for the lowest eigenvalue $\lambda_1$ (see Figure~\ref{figure:HO_consistent_radii}), in both cases less than $10^{-3}$ Ha. 
Note that there is a distinct difference in the PUM functions $\varphi_i \vartheta_i^m$ on even (e.g., $4\times4\times4$) and odd (e.g., $3\times3\times3$) covers. In the odd case, the potential center 
is located within the flat-top region of a single patch whereas for the even case $\vm{\tau}$ is contained in the overlap of eight neighboring patches. Thus, the results summarized in 
Figure~\ref{figure:HO_consistent_radii} show that our approach is also fully robust with respect to the relative position of $\vm{\tau}$.

\begin{figure}[tbp]
 \centering
 \begin{tikzpicture}[scale=0.7]
  \begin{loglogaxis}[
   xlabel={degrees of freedom},
   ylabel={time},
   grid=major,
   legend style={font=\scriptsize},legend pos=north west,
   legend entries={standard eigenproblem solver, generalized eigenproblem solver},
   ]
   \addplot[lime,mark=*,line width=0.8pt] table [x=dof, y=solve] {Data/RightTimings/Consistent_Timings_Small_Combined.dat};
   \addplot[teal,mark=x,line width=0.8pt] table [x=dof, y=solve] {Data/RightTimings/Lumped_Timings_Small_Combined.dat};
  \end{loglogaxis}
 \end{tikzpicture}
 \begin{tikzpicture}[scale=0.7]
  \begin{loglogaxis}[
   xlabel={degrees of freedom},
   ylabel={time},
   grid=major,
   legend style={font=\scriptsize},legend pos=north west,
   legend entries={standard eigenproblem solver, generalized eigenproblem solver},
   ]
   \addplot[lime,mark=*,line width=0.8pt] table [x=dof, y=solve] {Data/RightTimings/Consistent_Timings.dat};
   \addplot[teal,mark=x,line width=0.8pt] table [x=dof, y=solve] {Data/RightTimings/Lumped_Timings.dat};
  \end{loglogaxis}
 \end{tikzpicture}
 \caption{Wall clock time measured for the solution of the generalized and standard eigenproblems for the harmonic oscillator potential~\eqref{equation:harmonic_potential}, using the default SLEPc 
 eigenvalue solver, polynomials of degree $p=1,2,3$ and ten enrichments on an $n\times n\times n$ cover. All these data points are in one plot, sorted by number of DOFs 
 (left: $n=3,4,5,6,7,8$ on $16$ cores; right: $n=4,8,16,32$ on $64$ cores).}
 \label{figure:HO_Timings_largescale}
\end{figure}
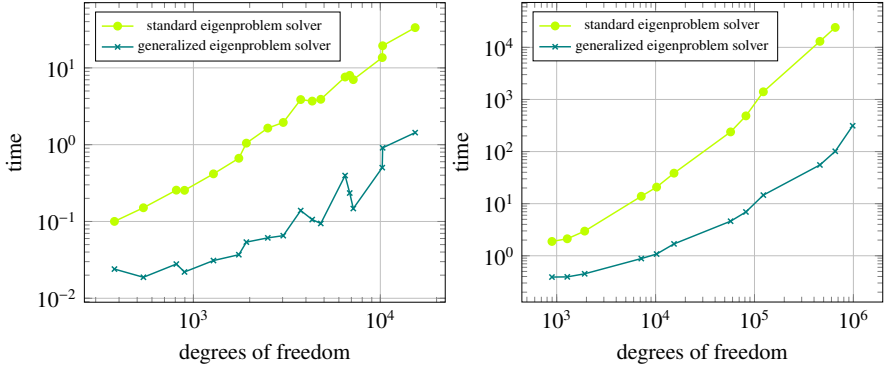
To properly evaluate the benefits of the lumped method, we first conduct the same experiment as above, only this time using a lumped overlap matrix and summarize the results in Figure 
\ref{figure:HO_lumped_radii}. We observe that the convergence behavior is practically identical to that of the consistent method, however, with a slightly larger absolute error.
Apart from being almost as accurate as the consistent method, using the lumped overlap matrix has its own benefits: First of all we can reduce the generalized eigenproblem 
\eqref{equation:generalized_eigenvalue_problem} to a standard eigenproblem as described in \sref{section:variational_mass_lumping}, which saves a lot of computational time, 
especially for larger problems (see~\fref{figure:HO_Timings_largescale}). 
In the context of self-consistent Kohn--Sham calculations this is especially important, as it enables the use of highly efficient algorithms to refine eigenvectors in each self-consistent-field 
iteration \cite{Zhou:2006,Banerjee:2016}. Computational time for the overall method is also again saved by using enrichment functions: while the computation
on a $16\times16\times16$ cover using quadratic polynomials without enrichments took $151.78$ seconds on $8$ cores, almost the same absolute error for the lowest eigenvalue $\lambda_1$ was achieved
by using a $3\times3\times3$ cover with quadratic polynomials, a single enrichment function for the lowest eigenvalue and an enrichment support radius of $4~\text{au}$ in $5.87$ seconds on $8$ cores.

\begin{table}[tbp]
\caption{Measured condition numbers with and without stabilization \eqref{equation:stabilityTransform} with respect to the local weighted $L^2$ inner product $L^2(\omega_i,\mathbb{C}; \varphi_i)$ 
for the consistent and the lumped overlap matrix obtained with $p=3$ on a $7\times7\times7$ uniform cover for increasing enrichment radius $r_e$ and ten enrichment functions per enriched patch.}
 \label{table:HO_condition}
 \centering
 \scalebox{0.8}{                                
 \begin{filecontents*}{Data/conditionL2/level3conditionmerged.dat}
           r             dof      condpre      condpost     lumpedpre     lumpedpost              
 0.000000000000000e+00  6860   4.419205e+03  7.453190e+02  5.793584e+01  1.000000e+00      
 5.000000000000000e-01  6870   1.862291e+11  8.724713e+02  7.670413e+10  1.000000e+00 
 1.000000000000000e+00  6930   3.596650e+11  2.107072e+03  8.543872e+10  1.000000e+00 
 1.500000000000000e+00  7150   4.531125e+11  3.383915e+03  9.361628e+10  1.000000e+00 
 2.000000000000000e+00  7590   6.213972e+11  5.377688e+03  1.237758e+11  1.000000e+00
 2.500000000000000e+00  8270   1.158239e+12  4.302366e+03  1.904394e+11  1.000000e+00
 3.000000000000000e+00  9090   2.395858e+12  4.600083e+03  4.082021e+11  1.000000e+00
 3.500000000000000e+00  9890   5.722176e+13  4.728283e+03  6.703891e+12  1.000000e+00
 4.000000000000000e+00  10210  2.901478e+14  4.699459e+03  2.099004e+13  1.000000e+00
\end{filecontents*}
\pgfplotstabletypeset[columns={r, dof, condpre, condpost, lumpedpre, lumpedpost},
                       columns/r/.style={column name=\textsc{$r_e$}, fixed, fixed zerofill, precision = 1},
                       columns/dof/.style={int detect, column name=\textsc{Dofs}},
                       columns/condpre/.style={column name = without \eqref{equation:stabilityTransform}, sci,sci zerofill, precision=0, string replace={-}{}},
                       columns/condpost/.style={column name = with \eqref{equation:stabilityTransform}, sci,sci zerofill, precision=0, string replace={-}{}},
                       columns/lumpedpre/.style={column name = without \eqref{equation:stabilityTransform}, sci,sci zerofill, precision=0, string replace={-}{}},
                       columns/lumpedpost/.style={column name = with \eqref{equation:stabilityTransform}, sci,sci zerofill, precision=0, string replace={-}{}},
                       empty cells with={--},
                       every head row/.style={before row={
                          \toprule
                          \multicolumn{1}{c}{} & \multicolumn{1}{c}{} & \multicolumn{2}{c}{consistent overlap matrix} & \multicolumn{2}{c}{lumped overlap matrix}\\
                          \midrule
                        },after row=\midrule},
                       every last row/.style={after row=\hline}
                       ]{Data/conditionL2/level3conditionmerged.dat} 
  }                      
 \end{table}

\begin{figure}[tbp]
 \centering
 \tikzset{every mark/.append style={scale=1.25}}
 \begin{tikzpicture}[scale=1.0]
  \begin{loglogaxis}[
   xlabel={degrees of freedom},
   ylabel={absolute error (Ha)},
   ymin=8e-08,
   grid=major,
   legend style={font=\tiny},legend pos=south west,
   legend entries={$p=1$, $p=2$, $p=3$, $(p=1) + e(\uparrow r_e)$, $(p=2) + e(\uparrow r_e)$, $(p=3) + e(\uparrow r_e)$},
   ]
   \addplot[teal,mark=triangle,line width=0.8pt] table [x=dof, y=e_0] {Data/gswithoutenrichlumped/Abs_errors_Harmonic_Oscillator_Lumped_poldeg1.dat};
   \addplot[lime,mark=square,line width=0.8pt] table [x=dof, y=e_0] {Data/gswithoutenrichlumped/Abs_errors_Harmonic_Oscillator_Lumped_poldeg2.dat};
   \addplot[magenta,mark=o,line width=0.8pt] table [x=dof, y=e_0] {Data/gswithoutenrichlumped/Abs_errors_Harmonic_Oscillator_Lumped_poldeg3.dat};
   \addplot[blue,mark=triangle*,line width=0.8pt] table [x=dof, y=e_0] {Data/groundstateincrradiusoddplumped/Abs_Errors_Harmonic_Oscillator_Lumped_enriched_poldeg1_incr_radius_l3_numenr1.dat};
   \addplot[green,mark=square*,line width=0.8pt] table [x=dof, y=e_0] {Data/groundstateincrradiusoddplumped/Abs_Errors_Harmonic_Oscillator_Lumped_enriched_poldeg2_incr_radius_l3_numenr1.dat};
   \addplot[red,mark=*,line width=0.8pt] table [x=dof, y=e_0] {Data/groundstateincrradiusoddplumped/Abs_Errors_Harmonic_Oscillator_Lumped_enriched_poldeg3_incr_radius_l3_numenr1.dat};
   \addplot[blue,mark=triangle*,line width=0.8pt] table [x=dof, y=e_0] {Data/groundstateincrradiusoddplumped/Abs_Errors_Harmonic_Oscillator_Lumped_enriched_poldeg1_incr_radius_l2_numenr1.dat};
   \addplot[green,mark=square*,line width=0.8pt] table [x=dof, y=e_0] {Data/groundstateincrradiusoddplumped/Abs_Errors_Harmonic_Oscillator_Lumped_enriched_poldeg2_incr_radius_l2_numenr1.dat};
   \addplot[red,mark=*,line width=0.8pt] table [x=dof, y=e_0] {Data/groundstateincrradiusoddplumped/Abs_Errors_Harmonic_Oscillator_Lumped_enriched_poldeg3_incr_radius_l2_numenr1.dat};
   \addplot[black,sharp plot,update limits=false,line width=1pt] coordinates {(10,1.00E-03) (10000000,1.00E-03)};
  \end{loglogaxis}
 \end{tikzpicture}\\[15pt]
 \begin{tikzpicture}[scale=1.0]
  \begin{loglogaxis}[
   xlabel={degrees of freedom},
   ylabel={absolute error (Ha)},
   ymin=8e-08,
   grid=major,
   legend style={font=\tiny},legend pos=south west,
   legend entries={$p=1$, $p=2$, $p=3$, $(p=1) + e(\uparrow r_e)$, $(p=2) + e(\uparrow r_e)$, $(p=3) + e(\uparrow r_e)$},
   ]
   \addplot[teal,mark=triangle,line width=0.8pt] table [x=dof, y=e_0] {Data/gswithoutenrichlumped/Abs_errors_Harmonic_Oscillator_Lumped_poldeg1.dat};
   \addplot[lime,mark=square,line width=0.8pt] table [x=dof, y=e_0] {Data/gswithoutenrichlumped/Abs_errors_Harmonic_Oscillator_Lumped_poldeg2.dat};
   \addplot[magenta,mark=o,line width=0.8pt] table [x=dof, y=e_0] {Data/gswithoutenrichlumped/Abs_errors_Harmonic_Oscillator_Lumped_poldeg3.dat};
   \addplot[blue,mark=triangle*,line width=0.8pt] table [x=dof, y=e_0] {Data/groundstateincrradius/Abs_Errors_Harmonic_Oscillator_Lumped_enriched_poldeg1_incr_radius_l3_numenr1.dat};
   \addplot[green,mark=square*,line width=0.8pt] table [x=dof, y=e_0] {Data/groundstateincrradius/Abs_Errors_Harmonic_Oscillator_Lumped_enriched_poldeg2_incr_radius_l3_numenr1.dat};
   \addplot[red,mark=*,line width=0.8pt] table [x=dof, y=e_0] {Data/groundstateincrradius/Abs_Errors_Harmonic_Oscillator_Lumped_enriched_poldeg3_incr_radius_l3_numenr1.dat};
   \addplot[blue,mark=triangle*,line width=0.8pt] table [x=dof, y=e_0] {Data/groundstateincrradius/Abs_Errors_Harmonic_Oscillator_Lumped_enriched_poldeg1_incr_radius_l2_numenr1.dat};
   \addplot[green,mark=square*,line width=0.8pt] table [x=dof, y=e_0] {Data/groundstateincrradius/Abs_Errors_Harmonic_Oscillator_Lumped_enriched_poldeg2_incr_radius_l2_numenr1.dat};
   \addplot[red,mark=*,line width=0.8pt] table [x=dof, y=e_0] {Data/groundstateincrradius/Abs_Errors_Harmonic_Oscillator_Lumped_enriched_poldeg3_incr_radius_l2_numenr1.dat};
   \addplot[black,sharp plot,update limits=false,line width=1pt] coordinates {(10,1.00E-03) (10000000,1.00E-03)};
  \end{loglogaxis}
 \end{tikzpicture}
 \caption{Convergence history of the lowest eigenvalue $\lambda_1$ for the harmonic oscillator potential~\eqref{equation:harmonic_potential} attained for different refinement schemes. We consider 
 a purely polynomial approximation ($p=1,2,3$) on a sequence of uniformly refined covers, and a refinement by increasing the enrichment radius (compare~\fref{figure:enrichment_radius}) with a single 
 enrichment function on a fixed uniform cover (top: $3\times3\times3$, $7\times7\times7$; bottom: $4\times4\times4$, $8\times8\times8$) that is labeled 
by $p=1,2,3 + e\uparrow r_e$. All results were obtained with 
 the lumped overlap matrix.}
\label{figure:HO_lumped_radii}
\end{figure}
\begin{figure}[tbp]
 \begin{subfigure}{0.33\textwidth}
  \centering
 \begin{tikzpicture}[scale=0.65]
  \begin{loglogaxis}[
   xlabel = {degrees of freedom},
   ylabel = {absolute error (Ha)},
   grid = major,
   legend style = {font=\tiny},
   legend entries = {$p= 1$, $p = 2$, $p= 3$},
   ]
   \addplot[blue,mark=triangle*,line width=0.8pt] table [x=dof, y=e_sum] {Data/LumpedConvergencebyalphatest/Abs_Errors_Harmonic_Oscillator_Lumped_enriched_poldeg1_stretch110_numenr10_fixres15.dat};
   \addplot[green,mark=square*,line width=0.8pt] table [x=dof, y=e_sum] {Data/LumpedConvergencebyalphatest/Abs_Errors_Harmonic_Oscillator_Lumped_enriched_poldeg2_stretch110_numenr10_fixres15.dat};
   \addplot[red,mark=*,line width=0.8pt] table [x=dof, y=e_sum] {Data/LumpedConvergencebyalphatest/Abs_Errors_Harmonic_Oscillator_Lumped_enriched_poldeg3_stretch110_numenr10_fixres15.dat};
   \addplot[black,sharp plot,update limits=false,line width=1pt] coordinates {(10,1.00E-03) (10000000,1.00E-03)};
   \addplot[black,sharp plot,update limits=false,line width=1pt, dashed] coordinates {(100,1.00E-07) (10000000,1.00E-07)};
  \end{loglogaxis}
 \end{tikzpicture}
 \subcaption{$\alpha=1.1$}
 \label{figure:HO_lumped_alpha1.1}
 \end{subfigure}
 \begin{subfigure}{0.33\textwidth}
 \centering
 \begin{tikzpicture}[scale=0.65]
  \begin{loglogaxis}[
   xlabel = {degrees of freedom},
   ylabel = {absolute error (Ha)},
   grid = major,
   legend style = {font=\tiny},
   legend entries = {$p= 1$, $p = 2$, $p= 3$},
   ]
   \addplot[blue,mark=triangle*,line width=0.8pt] table [x=dof, y=e_sum] {Data/LumpedConvergencebyalphatest/Abs_Errors_Harmonic_Oscillator_Lumped_enriched_poldeg1_stretch120_numenr10_fixres15.dat};
   \addplot[green,mark=square*,line width=0.8pt] table [x=dof, y=e_sum] {Data/LumpedConvergencebyalphatest/Abs_Errors_Harmonic_Oscillator_Lumped_enriched_poldeg2_stretch120_numenr10_fixres15.dat};
   \addplot[red,mark=*,line width=0.8pt] table [x=dof, y=e_sum] {Data/LumpedConvergencebyalphatest/Abs_Errors_Harmonic_Oscillator_Lumped_enriched_poldeg3_stretch120_numenr10_fixres15.dat};
   \addplot[black,sharp plot,update limits=false,line width=1pt] coordinates {(10,1.00E-03) (10000000,1.00E-03)};
   \addplot[black,sharp plot,update limits=false,line width=1pt, dashed] coordinates {(100,1.00E-07) (10000000,1.00E-07)};
  \end{loglogaxis}
 \end{tikzpicture}
 \subcaption{$\alpha=1.2$}
 \label{figure:HO_lumped_alpha1.2}
 \end{subfigure}
 \begin{subfigure}{0.33\textwidth}
 \centering
 \begin{tikzpicture}[scale=0.65]
  \begin{loglogaxis}[
   xlabel = {degrees of freedom},
   ylabel = {absolute error (Ha)},
   grid = major,
   legend style = {font=\tiny},
   legend entries = {$p= 1$, $p = 2$, $p= 3$},
   ]
   \addplot[blue,mark=triangle*,line width=0.8pt] table [x=dof, y=e_sum] {Data/LumpedConvergencebyalphatest/Abs_Errors_Harmonic_Oscillator_Lumped_enriched_poldeg1_stretch130_numenr10_fixres15.dat};
   \addplot[green,mark=square*,line width=0.8pt] table [x=dof, y=e_sum] {Data/LumpedConvergencebyalphatest/Abs_Errors_Harmonic_Oscillator_Lumped_enriched_poldeg2_stretch130_numenr10_fixres15.dat};
   \addplot[red,mark=*,line width=0.8pt] table [x=dof, y=e_sum] {Data/LumpedConvergencebyalphatest/Abs_Errors_Harmonic_Oscillator_Lumped_enriched_poldeg3_stretch130_numenr10_fixres15.dat};
   \addplot[black,sharp plot,update limits=false,line width=1pt] coordinates {(10,1.00E-03) (10000000,1.00E-03)};
   \addplot[black,sharp plot,update limits=false,line width=1pt, dashed] coordinates {(100,1.00E-07) (10000000,1.00E-07)};
  \end{loglogaxis}
 \end{tikzpicture}
 \subcaption{$\alpha=1.3$}
 \label{figure:HO_lumped_alpha1.3}
 \end{subfigure}
 \caption{Convergence history of the sum of the ten lowest eigenvalues for the harmonic oscillator potential~\eqref{equation:harmonic_potential} attained for 
 different values of $\alpha=1.1,1.2,1.3$ (left to right) on uniformly refined covers with ten enrichment functions employed on every patch $\omega_i$ with a lumped overlap matrix. The dashed line 
 indicates the accuracy of the employed reference solution.}
 \label{figure:HO_lumped_alpha}
\end{figure}
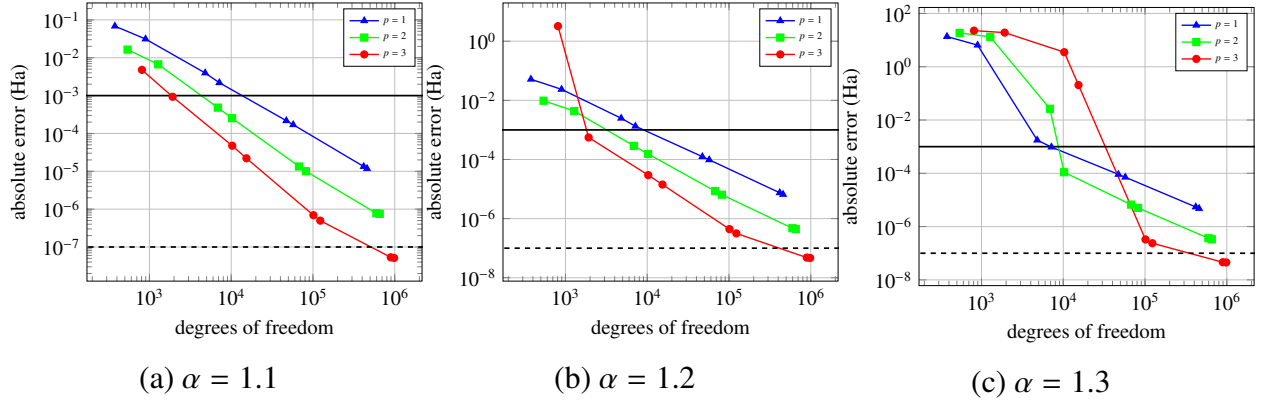

 Furthermore, we can overcome the problem of large condition numbers that arises in the PUFEM \cite{Pask:Sukumar:2017} using the stable transformation introduced in 
\sref{section:orbital_enrichment_and_stability}. The measured condition numbers for both consistent and lumped overlap matrices are displayed before and after 
applying the stable transformation \eqref{equation:stabilityTransform} in Table \ref{table:HO_condition}. Here we see that for both types of overlap matrices, the condition number deteriorates quickly with increasing 
enrichment support radius without our stabilization \eqref{equation:stabilityTransform}. The same observation was made in \cite{Pask:Sukumar:2017} for the PUFEM. However, after applying the 
stable transformation, we can dramatically reduce the condition number to the range of classical 
FE overlap matrices for the consistent case. Combining the stabilization \eqref{equation:stabilityTransform} with our lumping approach \eqref{equation:lumped-mass-matrix-pum} yields the optimal 
behavior $\bar{S}=\mathbb{I}$ so that \eqref{equation:generalized_eigenvalue_problem_lumped} becomes a well-conditioned standard eigenproblem. Thus, our flat-top PUM overcomes the two main drawbacks of the 
PUFEM and enables the stable use of enrichment functions on every patch and arbitrary potentials so that high-fidelity approximations can be attained with extremely small numbers of DOFs in a 
robust fashion. For example, the PUFEM employed in \cite{Sukumar:Pask:2009} yields an accuracy of $10^{-3}~\text{Ha}$ for this example with $948$ DOFs using enriched cubic FE where the condition number of the overlap matrix is of the order $O(10^{11})$ whereas our flat-top PUM with stability transform and lumping using a comparable setup provides an error of $5\cdot10^{-3}~\text{Ha}$ with $810$ DOFs and an optimal condition number of $1$ for the lumped overlap matrix.

Lastly, we examine the influence of the flat-top region on the eigenvalue convergence with the lumped overlap matrix. To this end, we consider various values of
the stretch factor~\eqref{equation:coverPatch} $\alpha=1.1,1.2,1.3$.~\fref{figure:HO_lumped_alpha} shows the convergence of the sum of
the ten lowest eigenvalues for different values of $\alpha$. For these tests, we enrich each of the ten lowest states on \emph{every} patch of our cover and use the stability transform
\eqref{equation:stabilityTransform}. From the depicted plots it becomes clear that lumping provides better results for smaller stretch factors $\alpha$. 
The absolute value of the error grows with increasing $\alpha$ as well as the pre-asymptotic range of the convergence behavior. For fine enough covers, i.e., in the asymptotic range, the convergence 
behavior is very much in agreement with the consistent scheme. These observations can be explained by the structure of the error introduced by the lumping scheme \cite{Schweitzer:2013*2}. Recall 
that our PUM degenerates to a DG approach with $\alpha=1$ and that the lumped mass matrix converges to the consistent matrix for $\alpha \rightarrow 1$. Thus, it is expected that the approximation 
quality of the computed eigenvalues will improve with smaller $\alpha$. The {\it a priori} determination of an optimal stretch factor $\alpha$ for a particular choice of the cover and employed local 
approximation spaces $V_i$, however, is the subject of current research.

\subsection{Gaussian potential} 
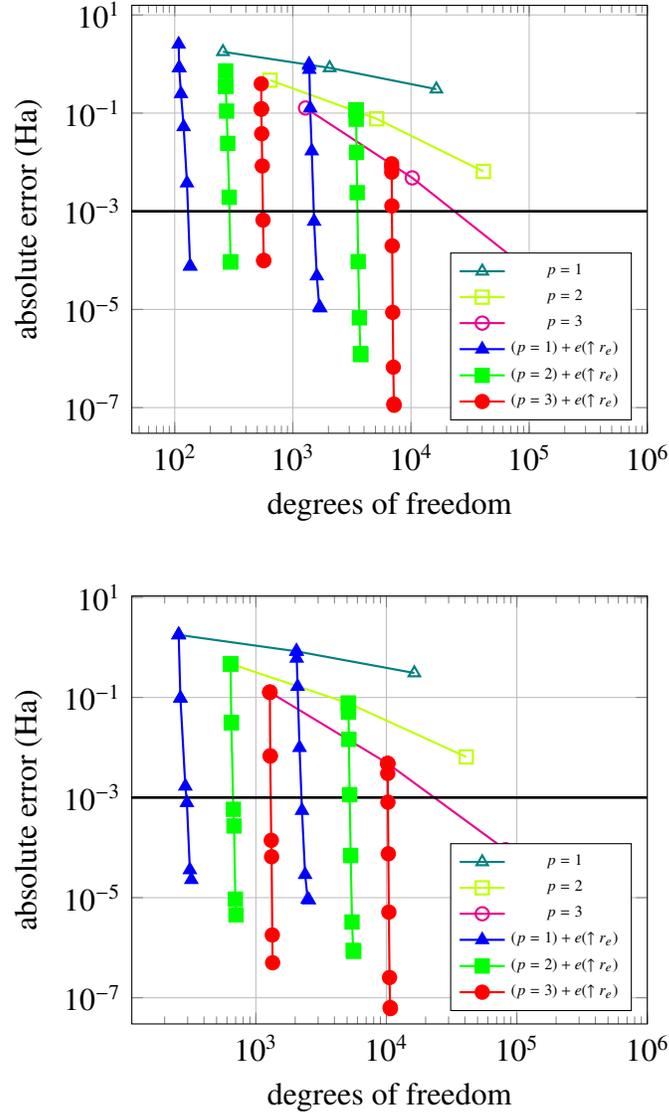
\begin{figure}[tbp]
 \centering
 \tikzset{every mark/.append style={scale=1.25}}
 \begin{tikzpicture}[scale=1.0]
  \begin{loglogaxis}[
   xlabel={degrees of freedom},
   ylabel={absolute error (Ha)},
   ymin=3e-08,
   xmax=10e05,
   grid=major,
   legend style={font=\tiny},legend pos=south east,
   legend entries={$p=1$, $p=2$, $p=3$, $(p=1) + e(\uparrow r_e)$, $(p=2) + e(\uparrow r_e)$, $(p=3) + e(\uparrow r_e)$},
   ]
   \addplot[teal,mark=triangle,line width=0.8pt] table [x=dof, y=e_0] {Data/GaussianWell/Abs_errors_Gaussian_Potential_Lumped_poldeg1.dat};
   \addplot[lime,mark=square,line width=0.8pt] table [x=dof, y=e_0] {Data/GaussianWell/Abs_errors_Gaussian_Potential_Lumped_poldeg2.dat};
   \addplot[magenta,mark=o,line width=0.8pt] table [x=dof, y=e_0] {Data/GaussianWell/Abs_errors_Gaussian_Potential_Lumped_poldeg3.dat};
   \addplot[blue,mark=triangle*,line width=0.8pt] table [x=dof, y=e_0] {Data/GaussianWell/Abs_Errors_Gaussian_Potential_Lumped_enriched_poldeg1_incr_radius_l2_numenr1_oddp.dat};
   \addplot[green,mark=square*,line width=0.8pt] table [x=dof, y=e_0] {Data/GaussianWell/Abs_Errors_Gaussian_Potential_Lumped_enriched_poldeg2_incr_radius_l2_numenr1_oddp.dat};
   \addplot[red,mark=*,line width=0.8pt] table [x=dof, y=e_0] {Data/GaussianWell/Abs_Errors_Gaussian_Potential_Lumped_enriched_poldeg3_incr_radius_l2_numenr1_oddp.dat};
   \addplot[blue,mark=triangle*,line width=0.8pt] table [x=dof, y=e_0] {Data/GaussianWell/Abs_Errors_Gaussian_Potential_Lumped_enriched_poldeg1_incr_radius_l3_numenr1_oddp.dat};
   \addplot[green,mark=square*,line width=0.8pt] table [x=dof, y=e_0] {Data/GaussianWell/Abs_Errors_Gaussian_Potential_Lumped_enriched_poldeg2_incr_radius_l3_numenr1_oddp.dat};
   \addplot[red,mark=*,line width=0.8pt] table [x=dof, y=e_0] {Data/GaussianWell/Abs_Errors_Gaussian_Potential_Lumped_enriched_poldeg3_incr_radius_l3_numenr1_oddp.dat};
   \addplot[black,sharp plot,update limits=false,line width=1pt] coordinates {(10,1.00E-03) (10000000,1.00E-03)};
  \end{loglogaxis}
 \end{tikzpicture}\\[15pt]
 \begin{tikzpicture}[scale=1.0]
  \begin{loglogaxis}[
   xlabel={degrees of freedom},
   ylabel={absolute error (Ha)},
   ymin=3e-08,
   xmax=10e05,
   grid=major,
   legend style={font=\tiny},legend pos=south east,
   legend entries={$p=1$, $p=2$, $p=3$, $(p=1) + e(\uparrow r_e)$, $(p=2) + e(\uparrow r_e)$, $(p=3) + e(\uparrow r_e)$},
   ]
   \addplot[teal,mark=triangle,line width=0.8pt] table [x=dof, y=e_0] {Data/GaussianWell/Abs_errors_Gaussian_Potential_Lumped_poldeg1.dat};
   \addplot[lime,mark=square,line width=0.8pt] table [x=dof, y=e_0] {Data/GaussianWell/Abs_errors_Gaussian_Potential_Lumped_poldeg2.dat};
   \addplot[magenta,mark=o,line width=0.8pt] table [x=dof, y=e_0] {Data/GaussianWell/Abs_errors_Gaussian_Potential_Lumped_poldeg3.dat};
   \addplot[blue,mark=triangle*,line width=0.8pt] table [x=dof, y=e_0] {Data/GaussianWell/Abs_Errors_Gaussian_Potential_Lumped_enriched_poldeg1_incr_radius_l2_numenr1.dat};
   \addplot[green,mark=square*,line width=0.8pt] table [x=dof, y=e_0] {Data/GaussianWell/Abs_Errors_Gaussian_Potential_Lumped_enriched_poldeg2_incr_radius_l2_numenr1.dat};
   \addplot[red,mark=*,line width=0.8pt] table [x=dof, y=e_0] {Data/GaussianWell/Abs_Errors_Gaussian_Potential_Lumped_enriched_poldeg3_incr_radius_l2_numenr1.dat};
   \addplot[blue,mark=triangle*,line width=0.8pt] table [x=dof, y=e_0] {Data/GaussianWell/Abs_Errors_Gaussian_Potential_Lumped_enriched_poldeg1_incr_radius_l3_numenr1.dat};
   \addplot[green,mark=square*,line width=0.8pt] table [x=dof, y=e_0] {Data/GaussianWell/Abs_Errors_Gaussian_Potential_Lumped_enriched_poldeg2_incr_radius_l3_numenr1.dat};
   \addplot[red,mark=*,line width=0.8pt] table [x=dof, y=e_0] {Data/GaussianWell/Abs_Errors_Gaussian_Potential_Lumped_enriched_poldeg3_incr_radius_l3_numenr1.dat};
   \addplot[black,sharp plot,update limits=false,line width=1pt] coordinates {(10,1.00E-03) (10000000,1.00E-03)};
  \end{loglogaxis}
 \end{tikzpicture}
 \caption{Convergence history of the lowest eigenvalue $\lambda_1$ for the periodic Gaussian potential \eqref{equation:gaussian_potential} using different 
 refinement schemes. We consider a purely polynomial approximation ($p=1,2,3$) on a sequence of uniformly refined covers, and a refinement by increasing the enrichment radius 
 (compare~\fref{figure:enrichment_radius}) with a single enrichment function on a fixed uniform cover (top: $3\times3\times3$, $7\times7\times7$; bottom: $4\times4\times4$, $8\times8\times8$) that is labeled by
 $p=1,2,3 + e\uparrow r_e$. All results were obtained with the lumped overlap matrix.}
 \label{figure:GW_lumped_radii}
\end{figure}

As our second benchmark problem, we consider a periodic Gaussian potential \cite{Sukumar:Pask:2009}. This external potential is defined via 
\begin{equation}
\label{equation:gaussian_potential}
 V(\vm{x}) = \sum_{\vm{R}} V_g(| \vm{x} - \vm{\tau} - \vm{R}|),
\end{equation}
with
\begin{equation}
 V_g(r) = -10 \, \text{exp}\left(- \frac{r^2}{2.25}\right),
\end{equation}
where we sum over the lattice translation vectors $\vm{R} = i_1 \vm{a}_1+i_2 \vm{a}_2+i_3 \vm{a}_3$, $i_d=-2,\ldots,2$ ($d=1,2,3$). The unit cell $\Omega$ and the potential center $\vm{\tau}$ are defined 
as in~\sref{subsection:HO}. For this experiment, we solve \eqref{equation:schroedinger_strong} subject to Bloch-periodic boundary conditions and choose 
$\vm{k} = (0.12, 0.23, 0.34)$ in reciprocal lattice coordinates.
We choose the same approximation spaces and stretch factor $\alpha=1.1$ as in~\sref{subsection:HO}. The problem-dependent radial part $R_{nl}$ of the infinite box eigenfunctions is multiplied 
by $h(r,10)$, i.e., $r_0=10~\text{au}$ in \eqref{equation:cut_off_function}. 
Also, as before, we are measuring the absolute error with respect to a cubic finite element reference solution computed on a $64\times64\times64$ mesh 
(with $\lambda_1^{\textrm{ref}}=-5.9605494576~\text{Ha}$ and $\sum_{i=1}^{10} \lambda_i^{\textrm{ref}} = -26.309704717~\text{Ha}$). For this example, we only show the convergence
results using a lumped overlap matrix and increasing the enrichment radius on a fixed cover (analogous to the situation studied in~\fref{figure:HO_lumped_radii}) to validate the method for 
Bloch-periodic boundary conditions.
The results obtained for the Gaussian potential are shown in~\fref{figure:GW_lumped_radii}. As for the harmonic oscillator example in \sref{subsection:HO}, we see the expected convergence
behavior of the non-enriched methods and a significant improvement of accuracy when increasing the number of patches that are enriched.
Again, the benefits of enrichment are striking: for this particular example, we only need as few as 135 DOFs (with linear polynomials on a $3\times3\times3$ cover) to achieve the desired accuracy
of $10^{-3}$ Ha for the lowest state.

\section{Concluding remarks}
\label{section:concluding_remarks}

In this paper, we addressed two key issues in quantum mechanical calculations using enriched partition of unity finite element methods: the need to solve a generalized rather than standard eigenvalue 
problem, and the ill-conditioning of the associated system matrices. To address these, we developed a stable and efficient orbital-enriched flat-top partition of unity method to solve the required 
Schr\"odinger equation subject to Bloch-periodic boundary conditions in a general parallelepiped domain. To this end, we employed a stable transformation and variational mass lumping in a 
flat-top partition of unity formulation. Compared to PUFEM approaches used previously, the present method yields well- (or even optimally-) conditioned system matrices and 
a standard eigenvalue problem rather than a generalized one, while maintaining accuracy and systematic convergence to benchmark results.

With the above key issues resolved, future work will focus on the incorporation of nonlocal pseudopotentials, as occur in practical calculations involving all but the lightest elements, 
systematic determination of maximal $\alpha$ consistent with chemical accuracy with variational mass lumping, and efficient numerical cubature. A corresponding implementation for the Poisson 
equation will then enable full Kohn--Sham calculations.

\section*{Acknowledgments}
This work was performed, in part, under the auspices of the U.S. Department of Energy by Lawrence Livermore National Laboratory under Contract DE-AC52-07NA27344.

\newpage


\bibliographystyle{elsarticle-num}
\bibliography{refs.bib}

\end{document}